\documentclass[twocolumn,aps,showpacs]{revtex4-1}
\usepackage[latin9]{inputenc}
\usepackage{xcolor}
\usepackage{pdfcolmk}
\usepackage{array}
\usepackage{float}
\usepackage{textcomp}
\usepackage{multirow}
\usepackage{amsmath}
\usepackage{amssymb}
\usepackage{graphicx}
\usepackage{setspace}
\PassOptionsToPackage{normalem}{ulem}
\usepackage{ulem}
\usepackage{subscript}

\makeatletter

\DeclareTextSymbolDefault{\textquotedbl}{T1}
\providecommand{\tabularnewline}{\\}
\providecolor{lyxadded}{rgb}{0,0,1}
\providecolor{lyxdeleted}{rgb}{1,0,0}

\DeclareRobustCommand{\lyxsout}[1]{\ifx\\#1\else\sout{#1}\fi}

\usepackage{color}

\def\NOT(#1,#2){\OneQubitGate(#1,#2){$X$}}

\makeatother

\begin{document}
\title{Optical spin initialization of spin-$\frac{3}{2}$ silicon vacancy
centers in 6H-SiC at room temperature}
\author{Harpreet Singh\textsuperscript{1}, Andrei N. Anisimov\textsuperscript{2},
\textcolor{black}{I. D. Breev$^{2}$,} Pavel G. Baranov\textsuperscript{2}
and Dieter Suter\textsuperscript{1}\\
 \textsuperscript{1}Fakultät Physik, Technische Universität Dortmund,\\
 D-44221 Dortmund, Germany. \textsuperscript{2}Ioffe Institute, St.
Petersburg 194021, Russia.}
\begin{abstract}
Silicon vacancies in silicon carbide have been proposed as an alternative
to nitrogen vacancy centers in diamonds for spintronics and quantum
technologies. An important precondition for these applications is
the initialization of the qubits into a specific quantum state. In
this work, we study the optical alignment of the spin 3/2 negatively
charged silicon vacancy in 6H-SiC. Using time-resolved optically detected
magnetic resonance, we coherently control the{\normalsize{} silicon
vacancy} spin ensemble and measure Rabi frequencies, spin-spin and
spin-lattice relaxation times of all three transitions. Then to study
the optical initialization process of the{\normalsize{} silicon vacancy}
spin ensemble,{\normalsize{} the vacancy} spin ensemble is prepared
in different ground states and optically excited. We describe a simple
rate equation model that can explain the observed behaviour and determine
the relevant rate constants.
\end{abstract}
\maketitle

\section{Introduction}

Silicon carbide (SiC) exists in many polytypes and hosts many interesting
vacancy centers, which have been shown to be useful for applications
in quantum technologies like sensing~\cite{falk-nature-13,widmann-nature-14,christle-nature-14,baranov2013,anisimov-aipa-2018,anisimov-sr-2018,koehl-nature-11}.Based on their spin in the ground state, these vacancy centers can be divided into two categories:
$S=1$ or $S=3/2$~\cite{riedel-prl-12,soykal-prb-16}.
Neutral divacancies, consisting of neighboring C and Si vacancies,
have spin 1. Four different types of divacancies in 4H-SiC have been
studied using optical and microwave techniques~\cite{koehl-nature-11}
similar to those used with nitrogen-vacancy qubits in diamond~\cite{Doherty:2013uq,suter-pnmrs-17,mr-2020-9}.
They can be efficiently polarized by optical irradiation and their
polarization can be transferred to $^{29}$Si nuclear spins, which
are strongly coupled to divacancies in 4H- and 6H-SiC~\cite{falk-prl-15}.
Coherent control of divacancy spins in 4H-SiC can be achieved even
at high temperature up to 600 K~\cite{yan-prap-18}. The spins of
neutral divacancies in SiC can sensitively detect both strain and
electric fields~\cite{falk-prl-14}, with higher sensitivity than
NV centers in diamond~\cite{falk-nature-13,falk-prl-14}.

Another type of vacancies consists of missing silicon atoms, i.e.,
silicon vacancies. If they capture an additional electron, they become
negatively charged silicon vacancies (V$_{Si}^{-}$) and have spin
3/2~\cite{riedel-prl-12,soykal-prb-16,soltamov-prl-12,baranov2013,biktagirov-prb-18}.
Several individually addressable silicon-vacancies have been identified
in different SiC polytypes. For example, the 6H-SiC hosts one hexagonal
site $h$ and two cubic sites ($k_{1}$ and $k_{2}$). V$_{Si}^{-}$
at $k_{1}$ and $k_{2}$ are called V$_{1}$ and V$_{3}$, respectively,
whereas V$_{Si}^{-}$ at the hexagonal sites $h$ are called V$_{2}$~\cite{sorman-prb-00,biktagirov-prb-18}.
Recently, it has been shown that $V_{Si}^{-}$ at hexagonal lattice
sites $h$ corresponds to V$_{1}$ and V$_{Si}^{-}$ at cubic lattice
sites $k_{1}$ and $k_{2}$ are V$_{3}$ and V$_{2}$, respectively~\cite{davidsson-apl-19}.
These negatively charged vacancies have zero phonon lines (ZPL) at 865 nm (V$ 1$), 887 nm (V$ 2$), and 908 nm (V$ 3$)~\cite{sorman-prb-00,biktagirov-prb-18,singh-prb-20}.
Optically induced alignment of the ground-state spin sublevels of
the V$_{Si}^{-}$ in 4H- and 6H-SiC has been demonstrated at room
temperature~\cite{soltamov-prl-12}. Coherent control of a single
silicon-vacancy spin and long spin coherence times have been reported~\cite{widmann-nature-14}.
V$_{Si}^{-}$ are relatively immune to electron-phonon interactions
and do not exhibit fast spin dephasing (spin coherence time $T_{2}=0.85$$\:$ms)~\cite{nagy-nc-19}.
Using a moderate magnetic field in combination with dynamic decoupling,
the spin coherence of the V$_{Si}^{-}$ spin ensemble in 4H-SiC with
natural isotopic abundance can be preserved over an unexpectedly long
time of \textgreater 20 ms$~$\cite{simin-prb-17}. Quantum microwave
emitters based on V$_{Si}^{-}$ in SiC at room temperature~\cite{kraus-nature-13}
can be enhanced via fabrication of Schottky barrier diodes and can
be modulated by almost 50\% by an external bias voltage~\cite{bathen-nature-15}.
Using all four levels, V$_{Si}^{-}$ can be used for absolute dc magnetometry
~\cite{soltamov-naturecom-19}.

In our previous work, we studied the temperature-dependent photoluminescence,
optically detected magnetic resonance (ODMR), and the relaxation times of the V$_{Si}^{-}$
spin ensemble in 6H-SiC at room temperature$~$\cite{singh-prb-20}.

In this work, we focus on the optical spin initialization of the V$_{1}$/V$_{3}$
in 6H-SiC, the spin relaxation and the dynamics of the intersystem-crossing.
Section~\ref{system} gives details of the optical pumping process.
Section~\ref{sec:expresult} explains the experimental setup for
continuous-wave (cw)- double-resonance and pulsed ODMR measurements.
Section~\ref{sec:Population-relaxation} describes the measurements
of the spin-lattice relaxation rates. Section \ref{sec:Optical-spin-alignment}
describes the dynamics of the optical spin alignment. Section~\ref{conc}
contains a brief discussion and concluding remarks.

\section{System}

\label{system}
The 6H-SiC sample we used is isotopically enriched in $^{28}$Si and $^{13}$C~\cite{singh-prb-20}.  The presence of 4.7 \% $^{13}$C reduces
the coherence time of the vacancy spin centers~\cite{yang-prb-14}.
Details of the sample preparation are given in Appendix$\:$A. In
our previous work~\cite{singh-prb-20}, PL spectra revealed the negative charged vacancies' zero phonon lines (ZPL) at 865 nm (V$ 1$), 887 nm (V$ 2$), and 908 nm (V$ 3$). 

\noindent 
\begin{figure}[H]
\centering \includegraphics{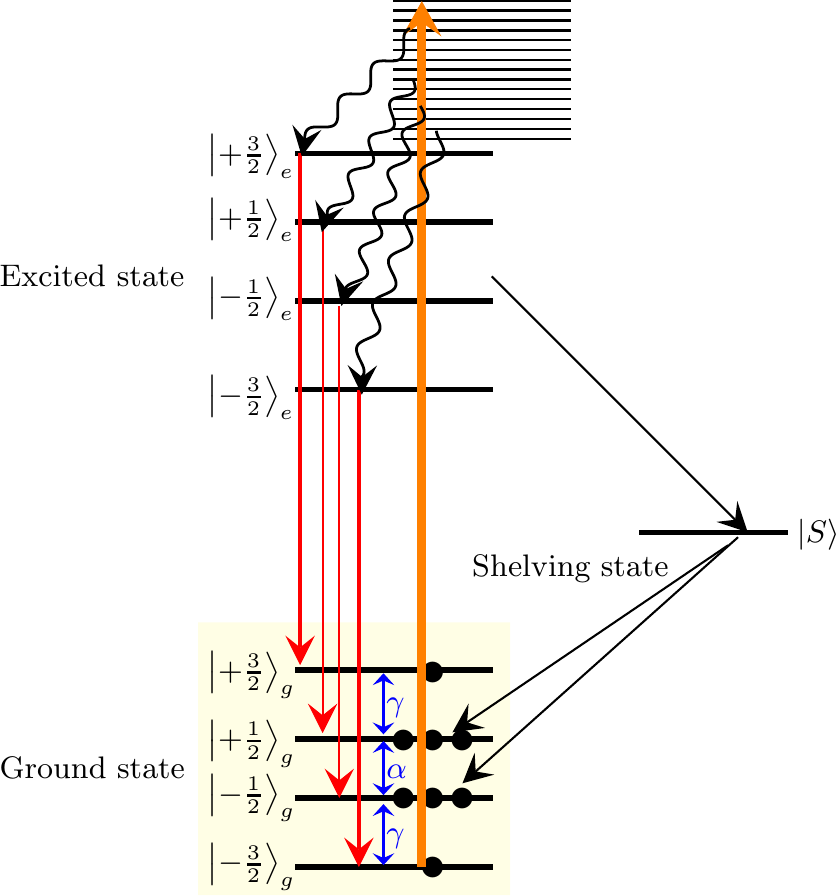} \caption{The ground, excited, and shelving states of the 6H-SiC V$_{1}$/V$_{3}$ form V$_{Si}^{-}$ are shown in this energy-level diagram. Red arrows indicate radiative transitions. An orange arrow for the non-resonant laser excitation. Black arrow shows the spin-dependent non-radiative transitions. The
states $\vert.\rangle_{g}$ represent ground states, $\vert.\rangle_{e}$
the excited states and $\vert S\rangle$ the shelving states.}
\label{energy_levels}
\end{figure}

The negatively charged V$_{1}$/V$_{3}$ type silicon vacancy center in 6H-SiC has
spin $S$ = 3/2~\cite{riedel-prl-12,soykal-prb-16}. Figure~\ref{energy_levels}
is the energy-level scheme in an external magnetic field.
The states $|.\rangle_{g}$  and $\vert.\rangle_{e}$
the electronically ground and excited states repectively~\cite{kraus-nature-13,biktagirov-prb-18,baranov-prb-11,fuchs-nature-15}. The shelving state $|S\rangle$ is an $S=1/2$ state.  This is essential
for the optical pumping process~\cite{baranov-prb-11} during which it gets populated by intersystem crossing (ISC).

The ground state spin Hamiltonian of the V$_{1}$/V$_{3}$ type defect is 
\begin{equation}
\mathcal{H}=D(S_{z}^{2}-\frac{5}{4}\hat{I})+g\mu_{B}\vec{B}\cdot\text{\ensuremath{\vec{S}}},\label{hamiltonian}
\end{equation}
 where  $g=2.0$ (electron $g$-factor), 
$2D=-28$ MHz (the zero field splitting)~\cite{biktagirov-prb-18,singh-prb-20},
 $\mu_{B}$ is the Bohr magneton,
$\hat{I}$ the unit operator,  $\vec{B}$ is the magnetic field strength,
and $\vec{S}$ is the four level electron spin operators.  We choose $z$-axis of our coordinate system such that it is parallel to the $c$-axis of the crystal (C$_{3}$ symmetry axis).

At ambient temperature, almost all four ground states have equal populations when the vacancy spin system is in thermal
equilibrium in the absence of optical pumping. The populations of the spin states
$\vert\pm\frac{1}{2}\rangle_{g}$ become large compared to those of
$\vert\pm\frac{3}{2}\rangle_{g}$, when the system is excited with a laser light, as shown schematically in Fig.~\ref{energy_levels}~\cite{baranov-prb-11,kraus-nature-13}. With the laser light illumination, the population of ground state goes to the excited states and with the spontaneous emission it come back to the ground states. During the excited state lifetime, which is $\sim$10 ns in the 6H
polytype, estimated from the linewidth of the excited state level
anti-crossing (LAC)$\;$\cite{astakhov2016spin} and $\sim$7.8 ns~
in 4H$\;$\cite{fuchs-nature-15,hain-jap-14}. Although, the spin system
can  go through intersystem-crossing (ISC) to the shelving states
$|S\rangle$~\cite{baranov-prb-11}. The measured time constant from
the excited state to $|S\rangle$ is $\sim$16.8 ns for V$_{Si}^{-}$
in 4H$~$\cite{fuchs-nature-15}. The system then returns to its initial state, with a preference for the states.
 $\vert\pm\frac{1}{2}\rangle$ is preferable to  $\vert\pm\frac{3}{2}\rangle$, with a time constant 150 ns for
V$_{Si}^{-}$ in 4H~\cite{fuchs-nature-15,riedel-prl-12,biktagirov-prb-18,soltamov-naturecom-19}.
The exact rates from the excited state to $|S\rangle$ and from $|S\rangle$
to the ground state have not been measured yet for V$_{Si}^{-}$ in
the 6H-SiC polytype, but they should be close to those in the 4H-SiC
polytype.

If the spins are not in thermal equilibrium and pumping stops, they
relax back to the thermal equilibrium state by spin-lattice relaxation,
as shown by the blue arrows in Fig.~\ref{energy_levels}. Here, $\gamma$
and $\alpha$ are the spin-lattice relaxation rates of the $\vert\pm\frac{3}{2}\rangle$$\longleftrightarrow$$\vert\pm\frac{1}{2}\rangle$
and $\vert+\frac{1}{2}\rangle$$\longleftrightarrow$$\vert-\frac{1}{2}\rangle$
transitions.

\section{Optically detected magnetic resonance}

\label{sec:expresult}

The ODMR technique is similar to the conventional electron spin resonance
(ESR) technique except for the additional optical pumping and the
detection part. In the ODMR technique, instead of measuring absorbed
microwave or radio frequency (RF) power, an optical signal is detected,
which may be photoluminescence (PL) or a transmitted or reflected
laser beam~\cite{Chen2003,Depinna-1982,langof-jpcb-02,mr-2020-9}.
The detailed description of the ODMR setup is given in Appendix~B.

\subsection{Continuous-wave ODMR}

\label{subsec:Continuous-wave-ODMR}

\begin{figure}
\includegraphics[scale=0.9]{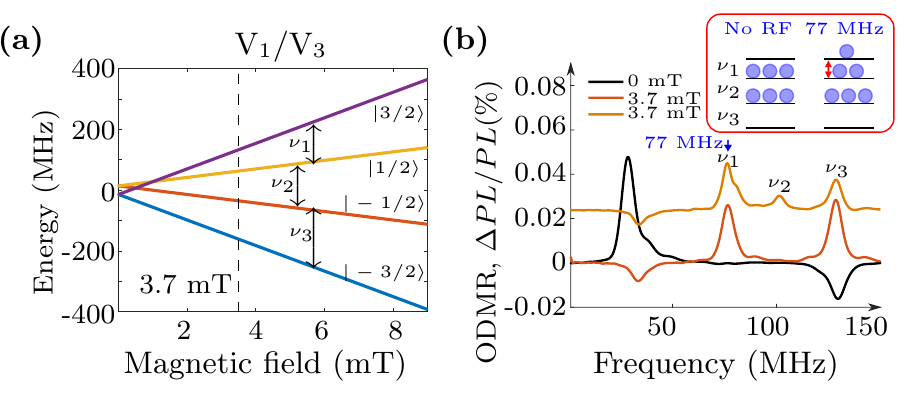}

\caption{(a) Energy levels of the V$_{1}$/V$_{3}$ vacancy in a magnetic field
$B$ $\parallel$ $c$-axis. (b) ODMR signals vs. frequency. The black
curve is the ODMR signal measured in the absence of a magnetic field,
the red curve is the ODMR signal in the 3.7 mT magnetic field, and
the orange curve is the ODMR with additional 77 MHz RF in the 3.7
mT magnetic field.}

\label{odmrplots}
\end{figure}

Figure~\ref{odmrplots}(b) depicts the ODMR signal measured in the
absence of a magnetic field by sweeping the direct digital synthesizer
(DDS) frequency as the black curve. Two peaks at 28 MHz and 128 MHz with opposite signs
are recorded. The PL signal is increase at 28 MHz RF and decrease at 128 MHz.
 Fig.~\ref{odmrplots}(a) shows energy levels of V$_{1}$/V$_{3}$ with magnetic field $B$ applied $ parallel$ $c$-axis, calculated using the given Hamiltonian in Eq.~\eqref{hamiltonian}.

The transition $\vert+3/2\rangle\leftrightarrow\vert+1/2\rangle$,
$\vert+1/2\rangle\leftrightarrow\vert-1/2\rangle$ and $\vert-3/2\rangle\leftrightarrow\vert-1/2\rangle$, are represented by arrows labeled with $\nu{}_{1}$, $\nu{}_{2}$ and $\nu{}_{3}$, respectively. With classical ODMR experiments, only two of the three
allowed transitions in the spin-3/2 system are observable, since the
$\pm1/2$ states have equal populations.

\noindent The ODMR signal recorded in a 3.7 mT magnetic field is plotted
as the red curve in Fig.~\ref{odmrplots}(b). The peak at 77 MHz
corresponds to the $\nu_{1}$ transition, and the peak at 129 MHz
corresponds to the $\nu_{3}$ transition. The negative peak around
the 30 MHz is due the V$_{2}$ type V$_{Si}^{-}$. Due to the equal
populations of the $\vert\pm1/2\rangle$ levels, no peak is visible
at the frequency of the $\nu_{2}$ transition. To observe this transition,
we added a second RF source to the setup, using it to selectively
change the populations. For these experiments, the pump frequency
was applied at frequency $\nu_{1}$ while the second device was scanned.
The output signals of both sources were combined with an RF combiner,
amplified and sent to the RF coils. The inset of Fig.~\ref{odmrplots}(b)
shows the modification of the populations by the pumping. Sweeping
the second DDS, we recorded the ODMR signal plotted as the orange
curve in Fig.~\ref{odmrplots}(b) where the $\nu_{2}$ transition
appears at 101 MHz. At room temperature, the PL from various types of vacancies cannot be separated$~$\cite{singh-prb-20}. As a result, the calculated PL contains contributions from other centers that are independent of magnetic resonance, resulting in a relatively small contrast.Further, the ODMR contrast of V$_{Si}^{-}$ depends on the ratio of these spontaneous and non-radiative transitions. So,
it would be interesting to find alternate laser excitation pathways
that provide higher ODMR contrast as well as spin polarization.

\subsection{Pulsed ODMR}

\label{subsec:Pulsed-ODMR}To measure rates and time constants, we
used time-resolved ODMR. To initialize  the V$_{Si}^{-}$ in all experiments, a laser pulse with a power of 100 mW and a length of 300 $mu$s was used,
i.e., populating the states $\vert\pm\frac{1}{2}\rangle$ more than
the states $\vert\pm\frac{3}{2}\rangle$. Following the polarization of the spin system, the system was subjected to a series of RF pulses, as detailed below. We applied a second laser pulse of duration 4 $\mu$s and
integrated the PL collected during the pulse to read the state of the spin system. To discard unnecessary background signals, the signal was summed 400 times and subtracted it from a reference experiment's 400 times summed signal. This procedure was carried out 20 times in total, with the average being taken each time~\cite{singh-prb-20}.

\noindent In the following, we assume that the population $\rho_{kk}$
of the $\pm3/2$ spin levels contributes a fraction $\Delta$ more
to the PL signal than the $\pm1/2$ spin levels~\cite{nagy-nc-19,carter-prb-15}.
The total PL signal $S$, measured with the second laser pulse, is
then

\begin{equation}
S=S_{+\frac{3}{2}}+S_{+\frac{1}{2}}+S_{-\frac{1}{2}}+S_{-\frac{3}{2}}
\end{equation}
with the contributions 

\begin{eqnarray*}
S_{+\frac{3}{2}} & = & (S_{0}+\Delta)\rho_{11}\\
S_{+\frac{1}{2}} & = & (S_{0}-\Delta)\rho_{22}\\
S_{-\frac{1}{2}} & = & (S_{0}-\Delta)\rho_{33}\\
S_{-\frac{3}{2}} & = & (S_{0}+\Delta)\rho_{44}
\end{eqnarray*}

\noindent from the populations of the different levels, where $S_{0}$
is the average signal contribution from each level. Taking into account
that the sum of the populations is =1, this can be further simplified
to

\begin{equation}
S=4S_{0}+\Delta(\rho_{11}-\rho_{22}-\rho_{33}+\rho_{44}).
\end{equation}

\begin{figure}[h]
\includegraphics{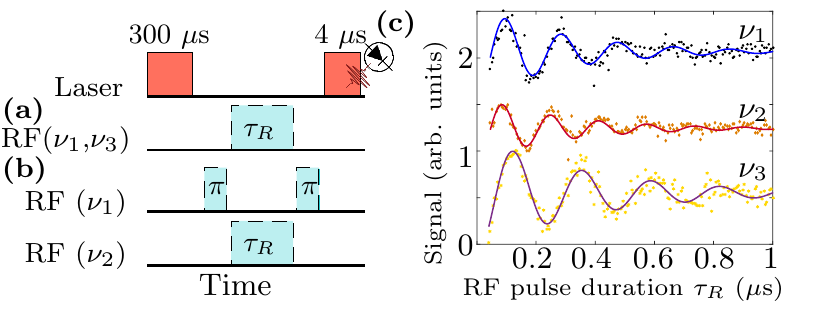} \caption{Pulse sequence for measuring Rabi oscillations of (a) $\nu_{1}$ and
$\nu_{3}$ transitions; (b) $\nu_{2}$ transition. The laser and RF pulses are represented by the red and light green boxes.  (c) Experimental Rabi
oscillations. The $y$-axis represents the PL signal change, while the $x$ axis represents the RF pulse length. }
\label{Rabi_all}
\end{figure}

We measured Rabi oscillations for the  $\nu_{1}$
and $\nu_{3}$ transitions of the V$ 1$/ V$ 3$  type V$_{Si}^{-}$,
using the pulse sequence scheme given in Fig.~\ref{Rabi_all}(a).
An RF pulse of $\tau_{R}$ duration was applied after the polarization laser pulse.
for the reference signal same experiment is repeated without RF Pulse. 

Rabi oscillations for the $\nu_{2}$ transition were measured using
the pulse sequence shown in Fig.~\ref{Rabi_all} (b). Two $\pi$
pulses with frequency $\nu_{1}$ were applied, and between them an
RF pulse with frequency $\nu_{2}$ and variable duration $\tau_{R}$.
The reference signal was obtained from an experiment without the RF
pulses. Figure~\ref{Rabi_all} (c) depicts the experimentally
 obtained data for the transitions at $\nu_{1}$, $\nu_{2}$ and $\nu_{3}$.
The following function was used to fit the experimental data

\begin{equation}
S_{RF}(\tau_{R})-S_{0}(\tau_{R})=A+B\text{ }cos(2\pi f_{R}\tau_{R}-\phi)e^{-\tau_{R}/T_{2}^{\text{\ensuremath{R}}}},\label{eq_rabi-1}
\end{equation}
here \textit{S}\textsubscript{\textit{RF}}\textit{($\tau_{R}$)
} is the PL signal measured with a duration \textit{ $\tau_{R}$
}, and \textit{S}\textsubscript{0}\textit{($\tau_{R}$)} is the reference signal measured with a delay $\tau_{R}$ and without the RF pulse. The Rabi frequencies
obtained with 20 W RF power and the measured dephasing times $T_{2}^{R}$
are given in Table~\ref{rabitable}. A plot of the Rabi frequencies
versus the square root of the RF power and a plot of the dephasing
times versus the Rabi frequencies of the transitions is given in Appendix
C.
\begin{center}
\begin{table}
\begin{centering}
\begin{tabular}{|c|c|c|}
\hline 
\multirow{1}{*}{Transition frequency} & $f_{R}$ (MHz) & $T_{2}^{R}$ (ns)\tabularnewline
\hline 
\hline 
$\nu_{1}$ (77 MHz) & $5.26\pm0.05$ & $299\pm30$\tabularnewline
\hline 
$\nu_{2}$ (101 MHz) & $6.14\pm0.05$ & $285\pm27$\tabularnewline
\hline 
$\nu_{3}$ (129 MHz) & $4.29\pm0.03$ & $381\pm29$\tabularnewline
\hline 
\end{tabular}
\par\end{centering}
\caption{Parameters of the Rabi oscillations for the different transitions
of V$_{1}$/ V$_{3}$ type V$_{Si}^{-}$.}

\label{rabitable}
\end{table}
\par\end{center}

\begin{figure}
\includegraphics{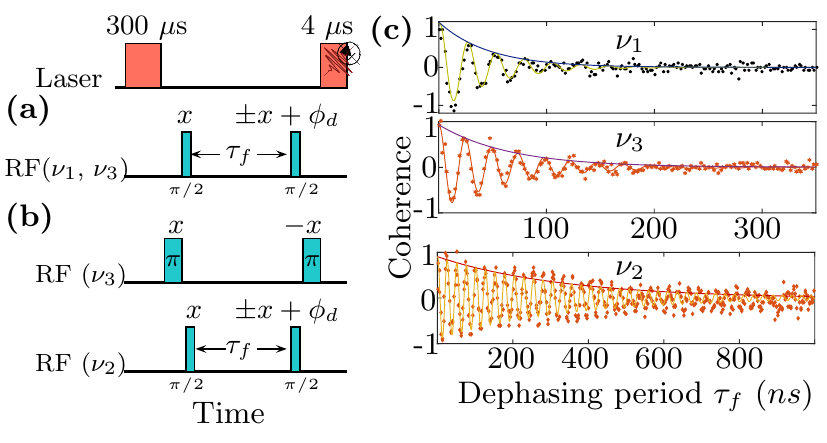}

\caption{Pulse sequence to measure the free induction decay of (a) $\nu_{1}$
and $\nu_{3}$ transitions; (b) $\nu_{2}$ transition. (c) FID signals
measured at the transitions $\nu_{1}$, $\nu_{3}$ and $\nu_{2}$.
The oscillation frequency is given by the detuning frequency of the
pulse sequence, which is 40 MHz in all 3 cases.}

\label{fid_all}
\end{figure}

Next, we performed free induction decay (FID) and spin-echo measurements
to measure the decay of coherence of all three transitions at room
temperature and in the 3.7 mT magnetic field. FID measurement include
coherence decay due to homogeneous and inhomogeneous interactions.
For FID measurements, we used the Ramsey scheme $\:$\cite{ramsey-pr-50}, in which the coherence was transformed into a population by the second $\pi/2$ RF pulse, after that, during the final laser pulse, the PL signal was read out.   Figure$\:$\ref{fid_all} (a) shows the pulse sequence scheme
for FID measurements of transitions $\nu_{1}$ and $\nu_{3}$ : First the system was initialized by the laser pulse, the coherence was created with the first $\pi/2$ RF
pulse of frequency $\nu_{1}$($\nu_{3}$),then allowed to evolve for a time $\tau_{f}$ and then
a $\pi/2$ RF pulse with phase $\phi_{d}=f_{det}\tau_{f}$ was applied
before the readout laser pulse. We took the difference between
two experiments, in which only difference is the second $\pi/2$ RF pulses have phases $\phi_{d}$ in one experiment
and $\pi+\phi_{d}$ in the other which suppress unwanted background
signals$\:$\cite{singh-prb-20}. The pulse sequence for the measurement
of the FID at $\nu_{2}$ is shown in Figure$\:$\ref{fid_all}(b).
It uses two additional $\pi$ pulses at frequency $\nu_{3}$ with
a phase difference of $\pi$, and 2 $\pi/2$ pulses at frequency $\nu_{2}$.
Figure$\:$\ref{fid_all} (c) shows the FIDs measured for $\nu_{1}$,
$\nu_{3}$ and $\nu_{2}$ respectively with detuning of $f_{det}$=
40 MHz, along with a fit function 
\begin{equation}
S^{FID}_{x+\phi_{d}}-S^{FID}_{-x+\phi_{d}}=A\:cos(2\pi f_{det}\tau_{f}+\phi)e^{-\tau_{f}/T_{2}^{*}}
\end{equation}
 where $S^{FID}_{x}+\phi_{d}(\tau_{f})$ and $S^{FID}_{-x}+\phi_{d}(\tau_{f})$
are the averaged PL signals measured with the second $\pi/2$ RF pulse with phase $\pm x+\phi_{d}$. The decay time $T_{2}^{*}$ is $46\pm6$ ns for transition
$\nu_{1}$, $333\pm40$ ns for transition $\nu_{2}$ and $66\pm8$
ns for transition $\nu_{3}$. The longer dephasing time for the transition
$\nu_{2}$ ($+1/2\leftrightarrow-1/2$) is expected as it is not affected
by the zero-field splitting$\:$\cite{EICKHOFF200469,soltamov-naturecom-19}
and therefore less affected by inhomogeneous broadening than the two
transitions $\nu_{1}$ and $\nu_{3}$.

\begin{figure}
\includegraphics{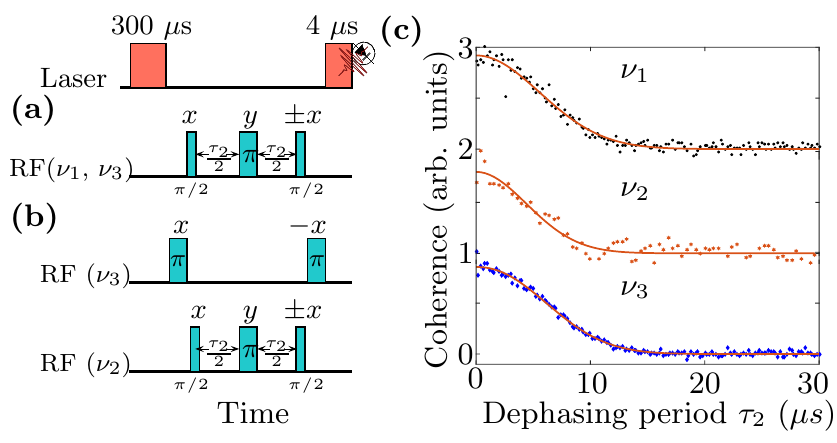}\caption{Pulse sequence for spin-echo experiments at (a) $\nu_{1}$ and $\nu_{3}$
transitions and (b) $\nu_{2}$ transition. (c) Spin-echo signals measured
at the transitions $\nu_{1}$, $\nu_{2}$, and $\nu_{3}$.}

\label{spin_echo}
\end{figure}

To measure the coherence decay due to the homogeneous interactions,
we performed the spin-echo experiment (Hahn echo)$\:$\cite{hahn-pr-50}.
Figure$\:$\ref{spin_echo}(a) shows the pulse sequence for the spin-echo
measurement of transitions $\nu_{1}$ and $\nu_{3}$, and Fig.$\:$\ref{spin_echo}(b)
shows for transition $\nu_{2}$. The pulse sequences used to measure
the spin-echo of all three transitions are similar to the corresponding
FID measurement sequences expect for an addition $\pi$ pulse in the
center between the two $\pi/2$ pulses. Figure$\:$\ref{spin_echo}(c),
plots the measured signal as a function of the dephasing period $\tau_{2}$
, along with a fit to the function
\begin{equation}
S_{x}-S_{-x}=Ae^{-(\tau_{2}/T_{2}^{\nu_{i}})^{n}},
\end{equation}
here the signals $S_{x}(\tau_{2})$ and $S_{-x}(\tau_{2})$ are measured
with the $\pm x$ phase $\pi/2$ pulse. The fitted parameters are $T_{2}^{\nu_{1}}$
= $7.9\pm0.2$$~$$\mu$s, $n$ = 2.23 for transition $\nu_{1}$,
$T_{2}^{\nu_{2}}$ = $6.2\pm0.3$$~$$\mu$s, $n$ = 1.97 for transition
$\nu_{2}$ and $T_{2}^{_{\nu_{3}}}$ = $8.2\pm0.3$$~$$\mu$s, $n$
= 2.17 for transition $\nu_{3}$ in the 3.7$~$$mT$
of an external magnetic field and at room temperature. The ratio $\langle T_{2}^{\nu_{1},\nu_{3}}\rangle/T_{2}^{\nu_{2}}=1.30\text{\ensuremath{\pm}}0.03$
agrees, within the experimental uncertainties, with the theoretical
value of 4/3 expected for relaxation by random magnetic fields coupling
to the electron spin dipole moment$\:$\cite{550}.

\section{Population relaxation}

\label{sec:Population-relaxation}

\subsection{Equation of motion}

\label{subsec:Equation-of-motion}

Uncontrolled interaction of a spin system with its environment causes dephasing (loss of coherence) and a return to the thermal equilibrium state when the system has been excited from its thermal equilibrium~\cite{abragam-book},
which is known as spin-lattice relaxation. In this process, energy
is exchanged between the system and its environment (the lattice).
As shown in Fig.~\ref{energy_levels}, due to energy exchange between
the V$_{Si}^{-}$ and their environment, the populations evolve towards
the equilibrium distribution with rates $\alpha$ and $\gamma$, where
$\alpha$ is the rate at which the $\vert\pm\frac{1}{2}\rangle_{g}$
spin levels equilibrate, and $\gamma$ is the rate between the $\vert\pm\frac{3}{2}\rangle_{g}\leftrightarrow\vert\pm\frac{1}{2}\rangle_{g}$
states . The time evolution of the four level system can thus be described
by the following equation:

\begin{eqnarray}
\frac{d}{dt}\vec{\rho} & = & \frac{1}{2}\left(\begin{array}{cccc}
-\gamma & \gamma\\
\gamma & -\alpha-\gamma & \alpha\\
 & \alpha & -\alpha-\gamma & \gamma\\
 &  & \gamma & -\gamma
\end{array}\right)\vec{\rho},\label{eq:t1}
\end{eqnarray}
where the population vector $\vec{\rho}$ contains the diagonal elements
$\rho_{ii}$ of the density operator.

The eigenvalues $\lambda_{i}$ and eigenvectors $\vec{u}_{i}$ for
Eq.~\eqref{eq:t1} are
\[
\vec{\lambda}=\left(\begin{array}{c}
0\\
-\gamma\\
-\frac{\alpha+\gamma+\xi}{2}\\
-\frac{\alpha+\gamma-\xi}{2}
\end{array}\right)
\]
and 
\[
\vec{u}_{i}=\left(\begin{array}{c}
1\\
1\\
1\\
1
\end{array}\right),\left(\begin{array}{c}
1\\
-1\\
-1\\
1
\end{array}\right),\left(\begin{array}{c}
-\gamma\\
\alpha+\xi\\
-\alpha-\xi\\
\gamma
\end{array}\right),\left(\begin{array}{c}
-\gamma\\
\alpha-\xi\\
-\alpha+\xi\\
\gamma
\end{array}\right)
\]
 respectively, where $\xi=\sqrt{\alpha^{2}+\gamma^{2}}$.

The solution of Eq.~\eqref{eq:t1} for an initial condition $\vec{\rho}(0)=(a,b,c,d)^{\top}$
is 
\begin{equation}
\vec{\rho}(t)=\frac{1}{4}\sum c_{i}e^{\lambda_{i}t}\vec{u}_{i}\label{eq:SolRelModel}
\end{equation}

where the weights

\begin{eqnarray*}
c_{1} & = & 1\\
c_{2} & = & a-b-c+d\\
c_{3} & = & \frac{(a-d)(\alpha-\xi)+\gamma(b-c)}{\gamma\xi}\\
c_{4} & = & -\frac{(a-d)(\alpha+\xi)+\gamma(b-c)}{\gamma\xi}
\end{eqnarray*}
are given by the initial conditions.

\subsection{Measurements}

\label{sec:relaxmeasure}

\begin{table}
\begin{centering}
\begin{tabular}{|c|c|c|}
\hline 
S.No & Initial state $\vec{\rho_{i}}(0)$ & Pulse sequence\tabularnewline
\hline 
\hline 
1 & $(0,0.5,0.5,0)^{\top}$ & $L$\tabularnewline
\hline 
2 & $(0.5,0.5,0,0)^{\top}$ & $L$$R^{\nu_{1}}$$R^{\nu_{2}}$\tabularnewline
\hline 
3 & $(0,0.5,0,0.5)^{\top}$ & $L$$R^{\nu_{3}}$\tabularnewline
\hline 
4 & $(0,0,0.5,0.5)^{\top}$ & $LR^{\nu_{3}}$$R^{\nu_{2}}$\tabularnewline
\hline 
\end{tabular}
\par\end{centering}
\caption{Pulse sequences used to prepare different initial states $\vec{\rho}_{i}(0)$,
where $L$ represents a laser pulse of duration 300 $\mu$s, and $R^{\nu_{i}}$
are RF pulses with frequency $\nu_{i}$ and flip angle $\pi$. Pulse
sequences 1 and 2 are used in Sec.$~$\ref{sec:Population-relaxation}
and 3 and 4 in Sec.$~$\ref{sec:Optical-spin-alignment}.}

\label{state_prep}
\end{table}

\begin{figure}[h]
\includegraphics{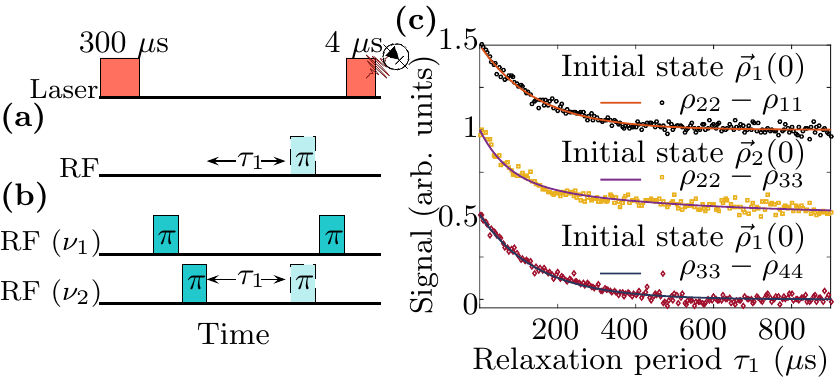}

\caption{Pulse sequences used to prepare specific initial states and measure
the population differences (a) $\text{\ensuremath{\rho{}_{22}}}(\tau_{1})-\text{\ensuremath{\rho{}_{11}}}(\tau_{1})$
and $\text{\ensuremath{\rho{}_{33}}}(\tau_{1})-\text{\ensuremath{\rho{}_{44}}}(\tau_{1})$,
and (b) $\text{\ensuremath{\rho{}_{22}}}(\tau_{1})-\text{\ensuremath{\rho{}_{33}}}(\tau_{1})$. the laser and RF pulses are represented by the red and light green boxes.
(c) Resulting experimental and calculated signals (from Eq.$~$\ref{eq:SolRelModel})
of the population differences with the delay $\tau_{1}$.}
\label{t1_all}
\end{figure}

High-quality measurements of the time dependence of individual populations
are difficult. We therefore measure differences $\rho_{ii}-\rho_{kk}$
between populations. We start with two particularly simple time-dependences,
which can be measured with the experiments shown in Fig.$~$\ref{t1_all}
(a) and (b). We first prepare an initial state with the populations
$\vec{\rho_{1}}(0)=(0,0.5,0.5,0)^{\top}$, so that $c_{1}=-c_{2}=1$
and $c_{3}=c_{4}=0$, and the expected time-dependence is

\begin{equation}
\vec{\rho}(t)=\frac{1}{4}(\vec{u}_{1}-e^{\lambda_{2}t}\vec{u}_{2})=\frac{1}{4}\left(\begin{array}{c}
1-e^{-\gamma t}\\
1+e^{-\gamma t}\\
1+e^{-\gamma t}\\
1-e^{-\gamma t}
\end{array}\right).\label{eq:t1_12_34}
\end{equation}

The pulse sequence for the preparation of the initial state $\vec{\rho_{_{1}}}(0)$
from the unpolarized state is shown in the first row of table \ref{state_prep}:
it consists of a laser pulse of duration 300 $\mu$s, which is long
enough to drive the system to a steady state. After the state preparation,
the system is allowed to relax for a time $\tau_{1}$. To read out
the final state, we apply an RF pulse with flip-angle $\pi$ and record
the PL during the measuring laser pulse. We subtract the result of
this experiment from a similar experiment where the RF pulse was omitted.
The resulting signal is proportional to the difference between the
populations that were exchanged by the $\pi$ pulse. If the $\pi$-pulse
is applied at frequency $\nu_{1}$, the signal is proportional to
$\text{\ensuremath{\rho{}_{22}}}(\tau_{1})-\text{\ensuremath{\rho{}_{11}}}(\tau_{1})$
and if it is applied at $\nu_{3}$, the signal is proportional to
$\text{\ensuremath{\rho{}_{33}}}(\tau_{1})-\text{\ensuremath{\rho{}_{44}}}(\tau_{1})$.

Figure$~$\ref{t1_all} (c) shows the resulting signals for $\text{\ensuremath{\rho{}_{22}}}(\tau_{1})-\text{\ensuremath{\rho{}_{11}}}(\tau_{1})$
and $\text{\ensuremath{\rho{}_{33}}}(\tau_{1})-\text{\ensuremath{\rho{}_{44}}}(\tau_{1})$
as a function of the delay $\tau_{1}$. We fit the theoretical signal
of Eq.$~$\eqref{eq:t1_12_34} to the experimental signal. From the
fits, we obtain the relaxation rate $\gamma=6.8\pm0.2$$~$ms$^{-1}$,
which corresponds to time constants $T_{1}^{12}$ = $T_{1}^{34}$=$1/\gamma$
=$146.2\pm3.6$$~$$\mu$s at room temperature.

To determine the second rate constant $\alpha$, a different initial
condition is needed. We chose $\vec{\rho_{2}}(0)=(0.5,0.5,0,0)^{\top}$
and measured the population difference $\rho_{22}-\rho_{33}$, which
we expect to depend on the relaxation delay $\tau_{1}$ as
\begin{eqnarray}
\rho_{22}(\tau_{1})-\rho_{33}(\tau_{1}) & = & \frac{\lambda_{3}(\alpha-\xi)e^{\lambda_{4}\tau_{1}}-\lambda_{4}(\alpha+\xi)e^{\lambda_{3}\tau_{1}}}{2\gamma\xi}.\label{eq:rho23}
\end{eqnarray}
The pulse sequence used to prepare the initial state $\vec{\rho_{2}}(0)$
from the thermal state is given in row 2 of Table$~\ref{state_prep}$
and in Figure$~$\ref{t1_all} (b) which also shows the sequence for
measuring the population difference $\rho_{22}-\rho_{33}$. For the
initial state preparation, the 300 $\mu$s laser pulse, and two RF
$\pi$ pulses were applied, one with frequency $\nu_{1}$ and a second
with frequency $\nu_{2}$. Then the system was allowed to evolve for
a time $\tau_{1}$ and another RF $\pi$-pulse with frequency $\nu_{1}$
and the signal measured using laser pulse. This experiment result
was subtracted from a reference experiment with an additional $\pi$
pulse of frequency $\nu_{2}$ after the delay $\tau_{1}$, as indicated
in Fig.$~$\ref{t1_all} (b) by the dashed rectangle. Fig.$~$\ref{t1_all}
(c) depict the obtained signals for $\text{\ensuremath{\rho{}_{22}}}(\tau_{1})-\text{\ensuremath{\rho{}_{33}}}(\tau_{1})$
with the delay $\tau_{1}$. The theoretical signal of
Eq.$~$\eqref{eq:rho23} was fitted to the experimental signals, using
the value of $\gamma$ determined before. From the fits, we obtained
the relaxation rate $\alpha=9.3\pm0.4$$~$ms$^{-1}$ and the relaxation
time $T_{1}^{23}$ = $1/\alpha$ =$107.3\pm4.9$$~$$\mu$s at room
temperature. The ratio $\alpha/\gamma=1.4\text{\ensuremath{\pm}}0.1$
agrees, within the experimental uncertainties, with the theoretical
value of 4/3 (the ratio of the squares of the corresponding transition
dipole moments) expected for relaxation by random magnetic fields
coupling to the electron spin dipole moment \cite{550}.

\section{Optical spin alignment}

\label{sec:Optical-spin-alignment}

\subsection{Experiments}

\begin{figure}
\includegraphics[scale=1.25]{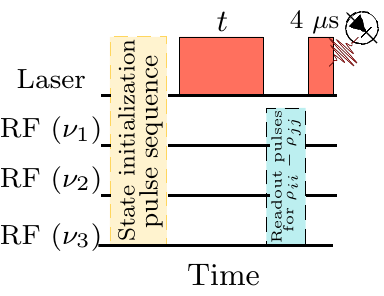}

\caption{Pulse sequence scheme used to prepare specific initial states and
measure the evolution of the population difference $\rho_{ii}-\rho_{jj}$
during optical pumping. Details of the initialization sequence are
given in Table II and for the readout sequence in table III.}

\label{gppscheme}
\end{figure}

\begin{figure}
\includegraphics[scale=1.08]{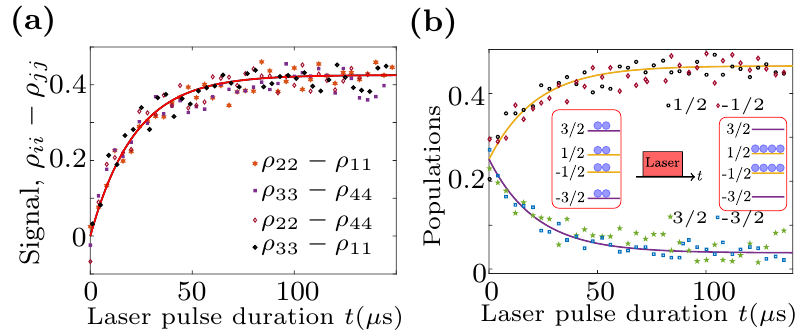}

\caption{(a) Population differences $\rho_{ii}-\rho_{jj}$, vs. laser pulse
duration $t$. Stars, squares, diamonds and circles represent the
experimentally measured signal for the population differences. The
red curve represents the analytical solution, obtained by solving
Eq.$~$\ref{eq:OptPumpSolution} when the system is initially unpolarized.
(b) Populations as a function of the laser pulse duration $t$, starting
from the unpolarized state.}

\label{thermalpop}
\end{figure}

Initialization of quantum registers to a specific state is one of
the primary requirements for the realization of any quantum device$~$\cite{divincenzo,Stolze:2008xy}.
The V$_{Si}^{-}$ spin ensemble can be initialized into the $\pm$1/2
spin states of the electronic ground-state by laser illumination$~$\cite{baranov-prb-11,kraus-nature-13,singh-prb-20}.
To determine the dynamics of this initialization process, we prepared
the spin ensemble in different initial states, applied a laser pulse
and again measured the populations $\vec{\rho}(t)$ as a function
of the duration of the laser pulse. Figure$~$\ref{gppscheme} shows
the pulse sequence used for preparing and measuring the population
differences $\rho_{ii}-\rho_{jj}$ during optical pumping. First,
we started with the unpolarized state $\vec{\rho}_{0}(0)=(1/4)(1,1,1,1)^{\top}$.
The laser pulse of duration $t$ was applied, followed by the RF sequence
given in Table$~$\ref{rfpulses} and the PL signal was measured during
the readout pulse. This PL signal was subtracted from a reference
PL signal measured by a similar experiment where no RF pulse was applied.
The difference signal is proportional to $\rho_{ii}-\rho_{jj}$. For
measuring the population differences $\rho_{22}-\rho_{11}$ and $\rho_{33}-\rho_{44}$,
the same RF pulse sequences were used as in Sec. \ref{sec:relaxmeasure}.
For measuring the population difference $\rho_{22}-\rho_{44}$, the
RF pulse sequence given in the third row of Table$~$\ref{rfpulses}
was used i.e., a $\text{\ensuremath{\pi}}$ pulse with frequency $\nu_{2}$
followed by a $\pi$ pulse at frequency $\nu_{3}$. For measuring
the population difference $\rho_{33}-\rho_{11}$, the RF pulse sequence
is given in the fourth row of Table$~$\ref{rfpulses}: A $\text{\ensuremath{\pi}}$
pulse at frequency $\nu_{2}$ is followed by another $\pi$ pulse
at frequency $\nu_{1}$. The experimental data $(S_{\rho_{ii}-\rho_{jj}}(t))$
were scaled by multiplying them with a constant factor $N$ such that
the signal for $\rho_{33}-\rho_{44}$ of the stationary state$\,$\eqref{eq:StationarySol}
prepared by a 300 $\mu$s laser pulse matches the theoretically expected
value of 0.37:

\begin{equation}
\rho_{ii}-\rho_{jj}(t)=NS_{\rho_{ii}-\rho_{jj}}(t).\label{eq:norm}
\end{equation}
 While the absolute scale of the signal is not important for the goal
of determining the rate constants, we use this scaling which fixes
the absolute values of the populations and allows a unique comparison
between the theoretical model and the experimental data. The resulting
normalized signals $\rho_{ii}-\rho_{jj}$ are shown in Fig.$~$\ref{thermalpop}(a).

\begin{table}
\begin{centering}
\begin{tabular}{|c|c|c|}
\hline 
S. No & Population difference & Pulse sequence\tabularnewline
\hline 
\hline 
1 & $\rho_{22}-\rho_{11}$ & $R^{\nu_{1}}$\tabularnewline
\hline 
2 & $\rho_{33}-\rho_{44}$ & $R^{\nu_{3}}$\tabularnewline
\hline 
3 & $\rho_{22}-\rho_{44}$ & $R^{\nu_{2}}R^{\nu_{3}}$\tabularnewline
\hline 
4 & $\rho_{33}-\rho_{11}$ & $R^{\nu_{2}}$$R^{\nu_{1}}$\tabularnewline
\hline 
\end{tabular}
\par\end{centering}
\caption{Readout pulse sequences used to measure different population differences.
$R^{\nu_{i}}$ are RF pulses with frequency $\nu_{i}$ and flip angle
$\pi$.}

\label{rfpulses}
\end{table}

When the V$_{Si}^{-}$ is excited with a non resonant laser, the shelving
states also populate the $m_{s}=\pm3/2$ states, and the resulting
spin polarization ($P$) is reduced$~$\cite{nagy-nc-19}. In this
more general case, the population vector can be written as

\begin{equation}
\vec{\sigma}=\frac{1-P}{4}\hat{I}+P\vec{\rho},
\end{equation}
and the polarization $P$ has been estimated as $\approx$80\%$\:$\cite{soltamov-prl-12}.
Here, $\vec{\sigma}$ represents the total ensemble's population vector,
and $\vec{\rho}=(\rho_{11},\rho_{22},\rho_{33},\rho_{44})^{\top}$
is the population vector of the spin-polarized sub-ensemble, which
can be manipulated by RF pulses. All four populations $\rho_{11}$,
$\rho_{22}$, $\rho_{33}$ and $\rho_{44}$ could be determined individually
from the four experiments described above and using the normalization
condition $\rho_{11}+\rho_{22}+\rho_{33}+\rho_{44}=1$.

Figure$~$\ref{thermalpop}(b) shows the evolution of the populations
of the spin states during the laser pulse. Starting from the unpolarized
state where all populations are $\rho_{ii}=1/4$, the populations
$\rho_{22}$ and $\rho_{33}$ grow to a limiting value of $\approx0.44$
while the populations $\rho_{11}$ and $\rho_{44}$ decrease to a
limiting value of $\approx0.06.$

\begin{figure}
\includegraphics{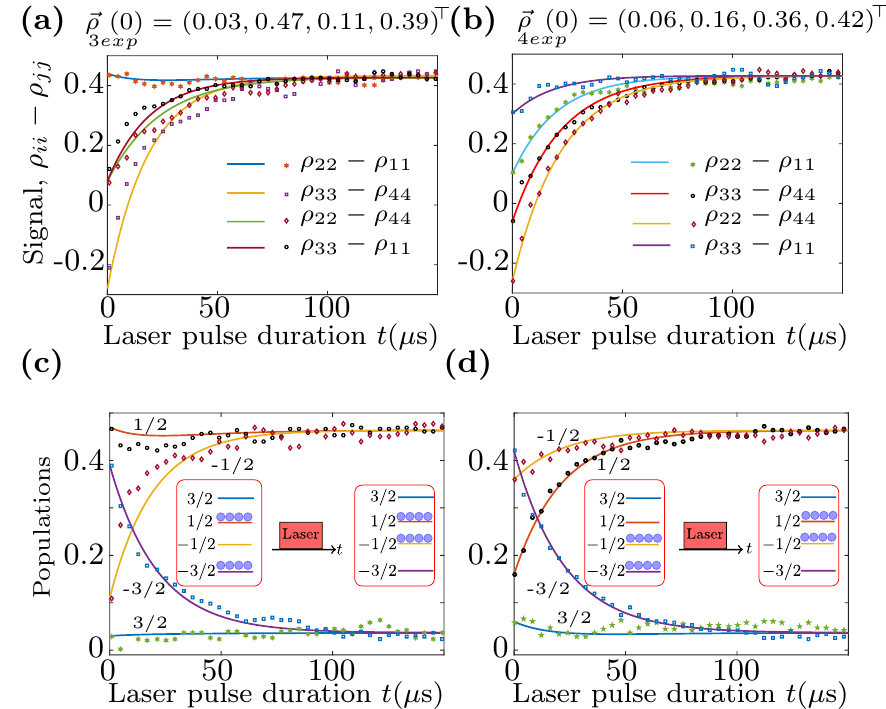}

\caption{a), b) : Population differences $\rho_{22}-\rho_{11}$, $\rho_{33}-\rho_{44}$,
$\rho_{22}-\rho_{44}$, and $\rho_{33}-\rho_{11}$ as a function of
the laser pulse duration $t$. The experimental data are represented
as points and the corresponding analytical solutions Eq. \eqref{eq:OptPumpSolution}
as curves. In (a), the initial condition was $\vec{\rho}_{3exp}(0)=(0.03,0.47,0.11,0.39)^{\top}$
and in (b) $\vec{\rho}_{4exp}(0)=(0.06,0.16,0.36,0.42)^{\top}$. c),
d) : Populations as a function of the laser pulse duration $t$, for
the initial state (c) $\vec{\rho}_{3exp}$ and (d) $\vec{\rho}_{4exp}$.}

\label{pp4scheme1a2}
\end{figure}

We repeated the experiment of Fig.$~$\ref{gppscheme}, with different
initial conditions: $\vec{\rho}_{3}(0)=(0,0.5,0,0.5)^{\top}$ and
$\vec{\rho_{4}}(0)=(0,0,0.5,0.5)^{\top}$. The pulse sequences used
for the preparation of these initial states are given in the third
and forth row of Table$~\ref{state_prep}$, respectively. The corresponding
results are shown in figures \ref{pp4scheme1a2}(a) and (b). From
the measured population differences, we reconstructed the time-dependence
of the populations, which are shown in Fig.$~$\ref{pp4scheme1a2}(c)
and (d), as a function of the laser pulse duration $t$. The experimentally
prepared initial states were $\vec{\rho}_{3exp}(0)=(0.03,0.47,0.11,0.39)^{\top}$
and $\vec{\rho}_{4exp}(0)=(0.06,0.16,0.36,0.42)^{\top}$. One of the
main reasons for the deviations of the experimentally prepared states
from the theoretical states is the limited laser intensity, which
results in incomplete polarisation. This could be improved by using
a tighter focus of the laser beam. Other causes are imperfections
in the RF pulses and relaxation.

\subsection{Model and rate constants}

\label{subsec:Model-and-rate}

The laser pulse transfers population from the spin-levels $\vert\pm\frac{3}{2}\rangle$
through the shelving state to $\vert\pm\frac{1}{2}\rangle$, as indicated
in Fig.$~$\ref{thermalpop}$~$(b). Figures$~$\ref{thermalpop}
and \ref{pp4scheme1a2} show that during the laser pulse, the populations
of the $\vert\pm\frac{1}{2}\rangle$ states increase to values close
to 0.5, while the populations of the $\vert\pm\frac{3}{2}\rangle$
states are strongly depleted.

Based on these experimental results and assuming that the lifetimes
in the excited state and the shelving states are short compared to
the pumping time, we use the following equations for modeling the
dynamics of the system:

\begin{singlespace}
\begin{align}
\frac{d}{dt}\vec{\rho} & =\nonumber \\
 & \frac{1}{2}\left(\begin{array}{cccc}
-\gamma-2\delta & \gamma\\
\gamma+\delta & -\alpha-\gamma-\delta & \alpha+\delta & \delta\\
\delta & \alpha+\delta & -\alpha-\gamma-\delta & \gamma+\delta\\
 &  & \gamma & -\gamma-2\delta
\end{array}\right)\vec{\rho}\label{eq:opticalmodel}
\end{align}
where $\delta$ is the rate at which population is pumped from the
states $\vert\pm3/2\rangle$ to $\vert\pm1/2\rangle$. The resulting
stationary state is
\begin{equation}
\vec{\rho}_{st}=\frac{\gamma}{4(\gamma+\delta)}\left(\begin{array}{c}
1\\
1\\
1\\
1
\end{array}\right)+\frac{\delta}{2(\gamma+\delta)}\left(\begin{array}{c}
0\\
1\\
1\\
0
\end{array}\right),\label{eq:StationarySol}
\end{equation}

\end{singlespace}

which approaches $(0,\frac{1}{2},\frac{1}{2},0)^{\top}$ for $\delta\gg\gamma$.

The eigenvalues and eigenvectors for Eq.~\eqref{eq:opticalmodel}
are

\[
\vec{\lambda}^{op}=\left(\begin{array}{c}
0\\
-\gamma-\delta\\
-\frac{\alpha+\gamma+\xi}{2}-\delta\\
-\frac{\alpha+\gamma-\xi}{2}-\delta
\end{array}\right)
\]

\begin{center}
$\begin{aligned}\vec{v}_{i}= & \left\{ \begin{aligned}\left(\begin{array}{c}
1\\
\frac{\gamma+2\delta}{\gamma}\\
\frac{\gamma+2\delta}{\gamma}\\
1
\end{array}\right),\left(\begin{array}{c}
1\\
-1\\
-1\\
1
\end{array}\right), & \left(\begin{array}{c}
-1\\
\frac{\alpha^{2}+\xi(\delta+\alpha)+\delta(\alpha-\gamma)}{\alpha\gamma+\delta(\alpha+\gamma-\xi)}\\
-\frac{\alpha+\xi}{\gamma}\\
1
\end{array}\right),\\
\left(\begin{array}{c}
-1\\
\frac{\alpha^{2}-\xi(\delta+\alpha)+\delta(\alpha-\gamma)}{\alpha\gamma+\delta(\alpha+\gamma+\xi)}\\
\frac{-\alpha+\xi}{\gamma}\\
1
\end{array}\right)
\end{aligned}
\right\} \end{aligned}
,$ respectively. The solution of Eq.~\eqref{eq:opticalmodel} for an
initial state $\vec{\rho}(0)=(a,b,c,d)^{\top}$ is
\par\end{center}

\begin{center}
\begin{equation}
\vec{\rho}(t)=\frac{1}{4}\sum p_{i}e^{\lambda_{i}^{op}t}\vec{v}_{i}\label{eq:OptPumpSolution}
\end{equation}
\par\end{center}

where

\begin{eqnarray*}
p_{1} & = & \frac{\gamma}{\gamma+\delta},\\
p_{2} & = & \frac{\gamma(a-b-c+d)+2\delta(a+d)}{\gamma+\delta},\\
p_{3} & = & \frac{\gamma(b-c)-(\xi-\alpha)(a-d)}{\xi},\\
p_{4} & = & \frac{\gamma(c-b)-(\xi+\alpha)(a-d)}{\xi}.
\end{eqnarray*}

The resulting expressions for the case where the initial state is
the depolarised state is given in Appendix$~$D. The calculated population
differences are plotted in Figures$~$\ref{thermalpop}$~$(a), \ref{pp4scheme1a2}$~$(a)
and (b), and the populations in Figures$~$\ref{thermalpop}$~$(b),
\ref{pp4scheme1a2}$~$(c) and (d). The best fits with the experimental
data, which were measured with a laser intensity $I=$ 622.64 W/cm\texttwosuperior{}
were obtained for the rate constant $\delta$ = $39\pm3$ $m$s$^{-1}$.
For $\alpha$ and $\gamma$, we used the values determined in section
\ref{sec:Population-relaxation}.

\begin{figure}
\begin{centering}
\includegraphics{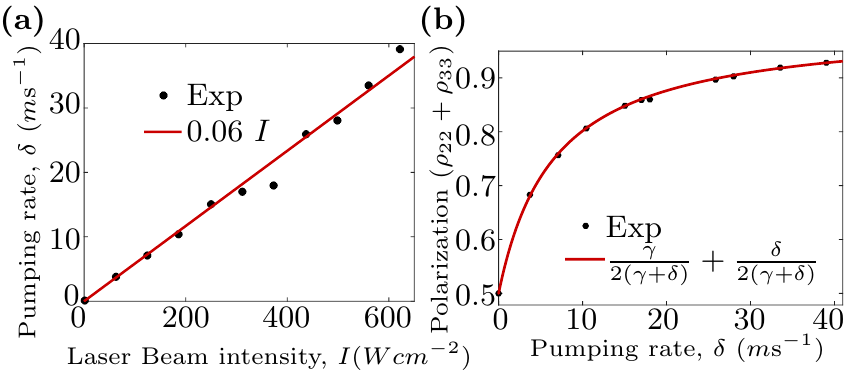}
\par\end{centering}
\caption{Plot of (a) pumping rates with the laser beam intensity, (b) polarization
with pumping rate.}

\label{laser_int}
\end{figure}

Taking additional data with different laser intensities, we found
that the pumping rate $\delta$ increases linearly with the intensity,
i.e., $\delta(\mathrm{ms^{-1}})=0.06\pm0.01$\ $I$$~$(W cm$^{-2}$),
as shown in Fig.$~$\ref{laser_int}$\:$(a). This indicates that
at the rate is limited by the population of the shelving state, which
is far from being saturated under our experimental conditions. Figure$~$\ref{laser_int}$\:$(b)
shows the spin polarization, measured as the sum of the $\pm$1/2
states after a laser pulse of 300 $\mu$s as a function of the pumping
rate $\delta$.

\section{Discussion and Conclusion}

\label{conc}

Silicon vacancy centers in SiC have shown promising results for quantum
sensing, single-photon emitters, and applications as light-matter
interfaces and other quantum technologies. In this work, we have demonstrated
the coherent control of all four levels of the V$_{1}$/ V$_{3}$
type V$_{Si}^{-}$. We measured the Rabi frequency of all three RF
transitions and the corresponding relaxation rates $\alpha$ (for
$\vert+1/2>$ $\leftrightarrow$ $\vert-1/2\rangle$ ) and $\gamma$
(for $\vert\pm3/2>$ $\leftrightarrow$ $\vert\pm1/2\rangle$ ) by
fitting the data with the proposed relaxation model. In general, the
relaxation of a spin 3/2 system can be described by three relaxation
modes with three time constants, which have been termed the spin dipole
($T_{p}$), quadrupole ($T_{d}$) and octupole ($T_{f}$)$\;$\cite{soltamov-naturecom-19,tarasenko-pssb-18,ramsay-prb-20,johan-cmr-03}.
For a dipole like perturbation the spin-relaxation times are $T_{p}$/3=$T_{d}$=2$T_{f}$,
$\alpha=(2/3)T_{d}^{-1}$ and $\gamma=(1/2)T_{d}^{-1}$ as discussed
theoretically in the case of a fluctuating magnetic field acting on
a V$_{Si}^{-}$ in SiC$\;$\cite{550,soltamov-naturecom-19}. The
ratio of the experimentally measured values of $\alpha$ and $\gamma$
is close to this theoretical value. It has recently been demonstrated
that for $V_{2}$ type V$_{Si}^{-}$ in 4H-SiC, this is not the case
due to mixing the octupole and dipole relaxation modes since a perturbation
with dipole symmetry cannot mix different order poles$\;$\cite{ramsay-prb-20}.

We also determined the dynamics of the optical initialization process
by preparing the system in three different initial states and measuring
the population dynamics during a laser pulse. The laser illumination
transfers the population from $\pm$3/2 to $\pm$1/2 spin states for
all three initial states. To interpret the resulting data, we proposed
a simple rate equation model for this process and were able to determine
all relevant rate constants from the experimental data. According
to Fig.$~$\ref{laser_int}$\:$(b), the spin polarisation of subensemble
$\rho$ approaches unity at intensitites \textgreater{} 200 W/cm$^{2}$.
As mentioned above, with the non-resonant laser excitation, the shelving
states also populate the \textpm{} 3/2 states, resulting in reduced
spin polarization of the total ensemble $\sigma$, which can be improved
by using resonant optical excitation$~$\cite{nagy-nc-19}. In conclusion,
we are confident that our results will contribute to a better understanding
of this fascinating system and open the way to more useful applications.
\begin{acknowledgments}
This research was financially supported by the Deutsche Forschungsgemeinschaft in the frame of the ICRC TRR 160 (Project No. C7) and the RFBR , project number 19-52-12058.
\end{acknowledgments}

\section*{Appendix A: sample preparation}
The experiments were carried out on the same sample as used in our previous work $~$\cite{singh-prb-20}. This sample is isotopically enriched in $^{28}$Si and $^{13}$C. As the source of a $^{28}$Si isotope, pre-prepared silicon available in the form of small pieces (1-3 mm) with an isotope
composition of 99.999\% $^{28}$Si was used. Carbon powder enriched to 15\% in $^{13}$C was used as the $^{13}$C source. The SiC crystal was grown in a laboratory. The SiC crystal was grown at a growth rate of $\approx$ 100$\mu$m/h on a (0001) Si face at a temperature of 2300°-2400°C in an argon atmosphere. Following the growth of the SiC crystal, the wafer was machined and cut. 
At room temperature, the crystal was irradiated with electrons of  an energy 2 MeV and with a dose of 10$^{18}$cm$^{-2}$ to create V$_{Si}^{-}$ centers.

\section*{Appendix B: ODMR setup}

\begin{figure}
\includegraphics[scale=0.85]{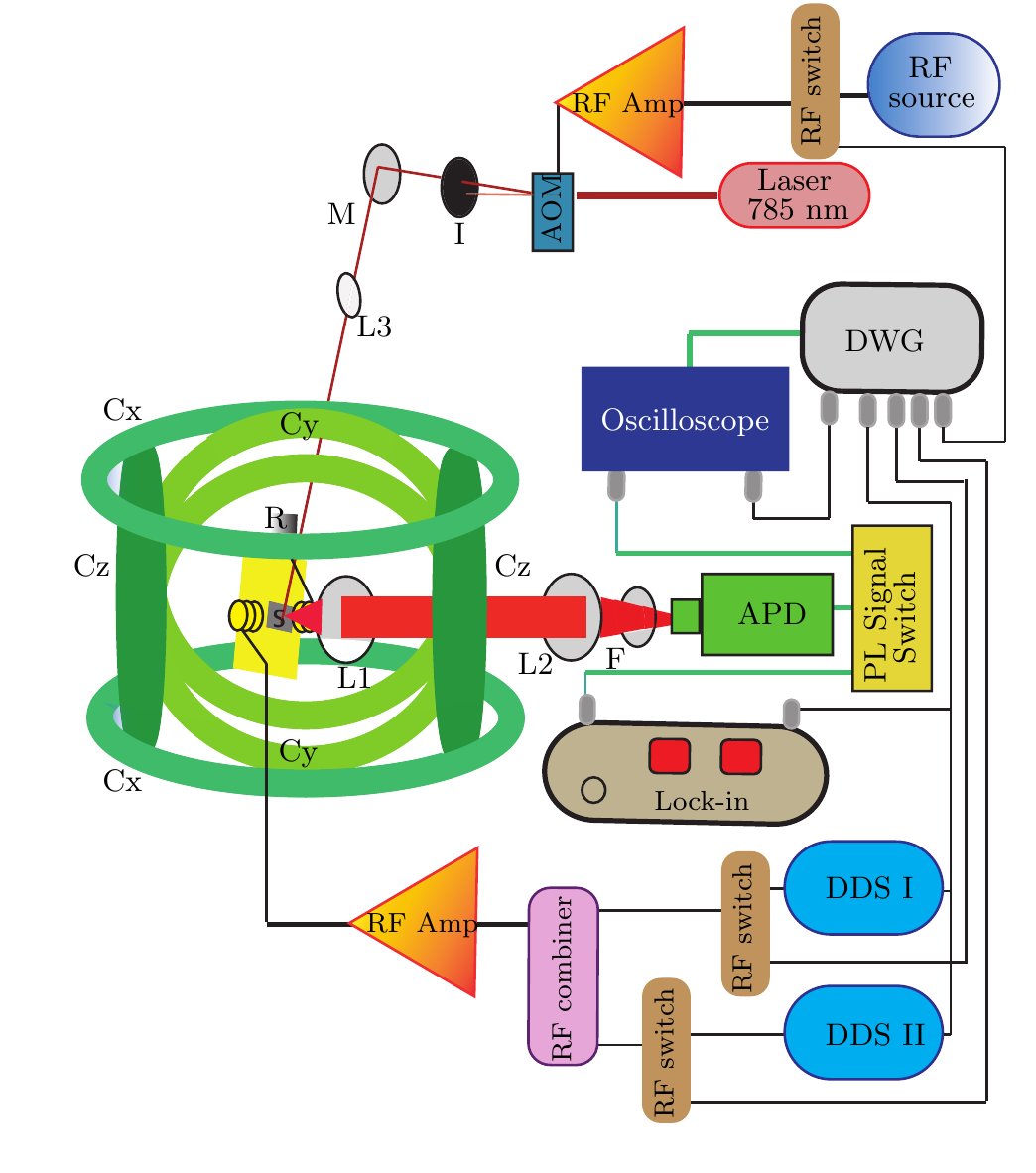} \caption{Experimental setup for measuring ODMR. The laser beam is represented by the red line from the laser.
The laser pulses are produced by the acousto-optical modulator (AOM). A long-pass filter, convex lenses, and a mirrors are defined by the ellipsoids named F, L, and M, respectively.  The RF is applied
using a three-turn Helmholtz coil-pair made of 100 $\mu$m diameter
copper wire placed perpendicular to the $c$-axis, in series with
a 50$\,$$\Omega$ resistor which is labeled
R. The SiC sample is the gray rectangle labeled S that is positioned between the RF coils.
Helmholtz coils for producing a static magnetic field in any direction are represented by the three orthogonal ring-pairs Cx, Cy, and Cz.
A digital word generator, an avalanche photodetector module, and direct digital synthesizers are represented by rounded rectangles labeled DWG, APD, and DDS, respectively. }
\label{odmr_setup}
\end{figure}

Figure~\ref{odmr_setup} shows the setup used for the cw- and time-resolved
ODMR measurements. A 785 nm laser diode (LD785-SE400) was used for the optical excitation source. An acousto-optical modulator (AOM; NEC model OD8813A) used for producing laser pulses. An RF switch (Mini-Circuits ZASWA-2-50DR+, dc-5 GHz) was used to produce the RF pulses. The Transistor-transistor logic pulses that control the timing were produced by a digital word generator (DWG; SpinCore PulseBlaster ESR-PRO PCI card). For applying the static magnetic field in any direction, we used three orthogonal Helmholtz coil-pairs. Currents up to 15A are delivered to the coils by current source (Servowatt, three-channel DCP-390/30).
An analog control voltage was used to control the currents individually. Analog Devices AD9915 direct digital synthesizer (DDS) provided the RF signal, which can generate signals up to 1 GHz. The RF pulses were created by an RF switch, amplified by a Mini-Circuits LZY-1 50 W amplifier, and applied to the SiC sample through a handmade Helmholtz-pair of RF coils with a diameter of 2.5 mm and three turns in each coil made from 100 $\mu$m diameter wire terminated with a 50 $\Omega$ resistor.  A convex lens (L3 in Fig.~\ref{odmr_setup}) with a focal length of 20 cm was used to focused a laser light on the sample. Two lenses L1 and L2, respectively, used to collect the PL. An avalanche photodiode (APD) module (C12703 series from Hamamatsu) used to record PL signal which was pass before through an 850 nm long-pass filter.  A USB card (PicoScope 2000 series) connected to a device (for pulsed measurements) or a lock-in amplifier (SRS model SR830 DSP) for cw-ODMR was used to record the signal from the APD. 
To measure the cw-ODMR signal, we used the setup shown in Fig.~\ref{odmr_setup}, with the RF switched on for continuous laser irradiation and the APD
connected to the lock-in amplifier. The DWG modulate the
amplitude of the RF and the APD signal was demodulated with
the lock-in amplifier whose reference signal was supplied by the DWG.

For the time-resolved ODMR, the APD shown in Fig.~\ref{odmr_setup}
was connected to the Picoscope USB card. The RF pulses were generated
using DDSes and RF switches and applied between the state initialization
and measurement.

\section*{Appendix C: POWER DEPENDENCE OF RABI OSCILLATIONS}

\subsection*{Rabi with RF Power}

\begin{figure}
\includegraphics{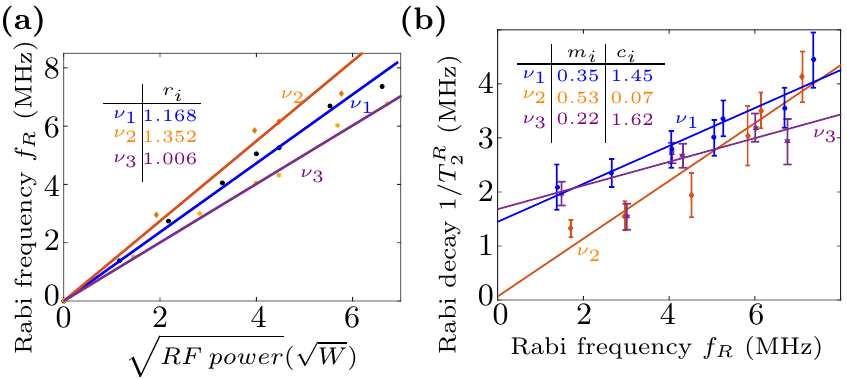}

\caption{(a) Plot of the experimentally measured Rabi frequency vs the square
root of RF Power. (b) Plot of the Rabi decay time $T_{2}^{R}$ of
different transitions vs the Rabi frequency.}

\label{rabi_rfp}
\end{figure}

\begin{table}
\begin{centering}
\begin{tabular}{|c|c|c|c|}
\hline 
\multirow{1}{*}{Transitions} & $r_{i}$ & $m_{i}$ & $c_{i}$\tabularnewline
\hline 
\hline 
$\nu_{1}$ & $1.168\pm0.062$ & $0.35\pm0.13$ & $1.45\pm0.66$\tabularnewline
\hline 
$\nu_{2}$ & $1.352\pm0.106$ & $0.53\pm0.24$ & $0.07\pm1.24$\tabularnewline
\hline 
$\nu_{3}$ & $1.006\pm0.041$ & $0.22\pm0.16$ & $1.62\pm0.84$\tabularnewline
\hline 
\end{tabular}
\par\end{centering}
\caption{Fitting parameters of Eqns \ref{eq:rabipower} and \ref{eq:rabi_relax}
for the different transitions of the V$_{1}$/ V$_{3}$ type V$_{Si}^{-}$.}

\label{rpr}
\end{table}

Figure$\:$\ref{rabi_rfp}(a) shows the measured Rabi frequencies
of the $\nu_{1}$, $\nu_{2}$ and $\nu_{3}$ transitions measured
versus the square root of the applied RF power. For fitting the experimental data we used the following function 
\begin{equation}
f_{R}(\sqrt{RF\:power})=r_{i}\sqrt{RF\:power}.\label{eq:rabipower}
\end{equation}

The fit parameters $r_{i}$ are given in the second column of Table$\:$\ref{rpr}
for the $\nu_{1}$, $\nu_{2}$, and $\nu_{3}$ transitions. The ratio
of the 3 Rabi frequencies is 1.173$\pm0.164$ : 2 : 1.149$\pm0.132$,
which is close to the theoretically expected value of $\sqrt{3}$
: $2$ : $\sqrt{3}$, i.e., the ratio of the transition dipole matrix
elements of a spin 3/2~\cite{ramsay-prb-20,Mizuochi-prb-02}. The
deviation appears to be related to reflections in the circuit that
feeds RF power to the sample, which generates standing waves and increases
with the RF frequency. In Fig.~\ref{rabi_rfp} (a), we measured the
forward RF power instead of actual RF current in the coil, which is
not easily accessible. Figure$\:$\ref{rabi_rfp}(b) shows the plot
of the Rabi decay times of the $\nu_{1}$, $\nu_{2}$ and $\nu_{3}$
transitions at different Rabi frequencies. The experimental data were
fitted to the function 
\begin{equation}
\frac{1}{T_{2}^{R}}(f_{R})=m_{i}f_{R}+c_{i}\label{eq:rabi_relax}
\end{equation}
 The fitted parameters $m_{i}$ and $c_{i}$ are given in the third
and fourth columns of Table$\:$\ref{rpr} for the $\nu_{1}$, $\nu_{2}$,
and $\nu_{3}$ transitions. The differences between the $T_{2}^{R}$
of $\nu_{1}$ and $\nu_{3}$ transitions are within the error margins
and appear not significant. At the lower Rabi frequency, the longer
dephasing time $T_{2}^{R}$ for the transition $\nu_{2}$ is expected
as it is less affected by inhomogeneous broadening than the two transitions
$\nu_{1}$ and $\nu_{3}$$\:$\cite{EICKHOFF200469,soltamov-naturecom-19}.

\section*{Appendix D: Time dependences for specific initial conditions}

This section provides two specific solutions for the population dynamics
without and with the laser field, for 2 different initial conditions.
For the initial state $\vec{\rho_{2}}(0)=\text{\ensuremath{\frac{1}{2}}}(1,1,0,0)^{\top}$,
the solution of the relaxation dynamics Eq.~\eqref{eq:SolRelModel}
is
\begin{center}
\begin{eqnarray*}
\rho_{11}(t) & = & \frac{\lambda_{3}e^{\lambda_{4}t}+\lambda_{4}e^{\lambda_{3}t}+\xi}{4\xi};\\
\rho_{22}(t) & = & \frac{\lambda_{3}(\alpha-\xi)e^{\lambda_{4}t}-\lambda_{4}(\alpha+\xi)e^{\lambda_{3}t}+\gamma\xi}{4\gamma\xi};\\
\rho_{33}(t) & = & \frac{\lambda_{3}(-\alpha+\xi)e^{\lambda_{4}t}+\lambda_{4}(\alpha+\xi)e^{\lambda_{3}t}+\gamma\xi}{4\gamma\xi};\\
\rho_{44}(t) & = & \frac{-\lambda_{3}e^{\lambda_{4}t}+\lambda_{4}e^{\lambda_{3}t}+\xi}{4\xi}.
\end{eqnarray*}
\par\end{center}

For the initial state $\vec{\rho_{0}}(0)=\frac{1}{4}(1,1,1,1)^{\top}$,
the solution of of the optical pumping dynamics Eq.~\eqref{eq:OptPumpSolution}
is

\begin{eqnarray}
\rho_{11} & = & \rho_{44}=-\frac{\gamma+\delta e^{\lambda_{2}^{op}t}}{4\lambda_{2}^{op}},\label{eq:rho44thermalmodel}\\
\rho_{22} & = & \rho_{33}=-\frac{\gamma+2\delta-\delta e^{\lambda_{2}^{op}t}}{4\lambda_{2}^{op}}.\label{eq:rho33thermal}
\end{eqnarray}


\begin{thebibliography}{45}%
\makeatletter
\providecommand \@ifxundefined [1]{%
 \@ifx{#1\undefined}
}%
\providecommand \@ifnum [1]{%
 \ifnum #1\expandafter \@firstoftwo
 \else \expandafter \@secondoftwo
 \fi
}%
\providecommand \@ifx [1]{%
 \ifx #1\expandafter \@firstoftwo
 \else \expandafter \@secondoftwo
 \fi
}%
\providecommand \natexlab [1]{#1}%
\providecommand \enquote  [1]{``#1''}%
\providecommand \bibnamefont  [1]{#1}%
\providecommand \bibfnamefont [1]{#1}%
\providecommand \citenamefont [1]{#1}%
\providecommand \href@noop [0]{\@secondoftwo}%
\providecommand \href [0]{\begingroup \@sanitize@url \@href}%
\providecommand \@href[1]{\@@startlink{#1}\@@href}%
\providecommand \@@href[1]{\endgroup#1\@@endlink}%
\providecommand \@sanitize@url [0]{\catcode `\\12\catcode `\$12\catcode
  `\&12\catcode `\#12\catcode `\^12\catcode `\_12\catcode `\%12\relax}%
\providecommand \@@startlink[1]{}%
\providecommand \@@endlink[0]{}%
\providecommand \url  [0]{\begingroup\@sanitize@url \@url }%
\providecommand \@url [1]{\endgroup\@href {#1}{\urlprefix }}%
\providecommand \urlprefix  [0]{URL }%
\providecommand \Eprint [0]{\href }%
\providecommand \doibase [0]{http://dx.doi.org/}%
\providecommand \selectlanguage [0]{\@gobble}%
\providecommand \bibinfo  [0]{\@secondoftwo}%
\providecommand \bibfield  [0]{\@secondoftwo}%
\providecommand \translation [1]{[#1]}%
\providecommand \BibitemOpen [0]{}%
\providecommand \bibitemStop [0]{}%
\providecommand \bibitemNoStop [0]{.\EOS\space}%
\providecommand \EOS [0]{\spacefactor3000\relax}%
\providecommand \BibitemShut  [1]{\csname bibitem#1\endcsname}%
\let\auto@bib@innerbib\@empty
\bibitem [{\citenamefont {Falk}\ \emph {et~al.}(2013)\citenamefont {Falk},
  \citenamefont {Buckley}, \citenamefont {Calusine}, \citenamefont {Koehl},
  \citenamefont {Dobrovitski}, \citenamefont {Politi}, \citenamefont {Zorman},
  \citenamefont {Feng},\ and\ \citenamefont {Awschalom}}]{falk-nature-13}%
  \BibitemOpen
  \bibfield  {author} {\bibinfo {author} {\bibfnamefont {A.~L.}\ \bibnamefont
  {Falk}}, \bibinfo {author} {\bibfnamefont {B.~B.}\ \bibnamefont {Buckley}},
  \bibinfo {author} {\bibfnamefont {G.}~\bibnamefont {Calusine}}, \bibinfo
  {author} {\bibfnamefont {W.~F.}\ \bibnamefont {Koehl}}, \bibinfo {author}
  {\bibfnamefont {V.~V.}\ \bibnamefont {Dobrovitski}}, \bibinfo {author}
  {\bibfnamefont {A.}~\bibnamefont {Politi}}, \bibinfo {author} {\bibfnamefont
  {C.~A.}\ \bibnamefont {Zorman}}, \bibinfo {author} {\bibfnamefont {P.~X.-L.}\
  \bibnamefont {Feng}}, \ and\ \bibinfo {author} {\bibfnamefont {D.~D.}\
  \bibnamefont {Awschalom}},\ }\href {\doibase 10.1038/ncomms2854} {\bibfield
  {journal} {\bibinfo  {journal} {Nature communications}\ }\textbf {\bibinfo
  {volume} {4}},\ \bibinfo {pages} {1819} (\bibinfo {year} {2013})}\BibitemShut
  {NoStop}%
\bibitem [{\citenamefont {Widmann}\ \emph {et~al.}(2015)\citenamefont
  {Widmann}, \citenamefont {Lee}, \citenamefont {Rendler}, \citenamefont {Son},
  \citenamefont {Fedder}, \citenamefont {Paik}, \citenamefont {Yang},
  \citenamefont {Zhao}, \citenamefont {Yang}, \citenamefont {Booker} \emph
  {et~al.}}]{widmann-nature-14}%
  \BibitemOpen
  \bibfield  {author} {\bibinfo {author} {\bibfnamefont {M.}~\bibnamefont
  {Widmann}}, \bibinfo {author} {\bibfnamefont {S.-Y.}\ \bibnamefont {Lee}},
  \bibinfo {author} {\bibfnamefont {T.}~\bibnamefont {Rendler}}, \bibinfo
  {author} {\bibfnamefont {N.~T.}\ \bibnamefont {Son}}, \bibinfo {author}
  {\bibfnamefont {H.}~\bibnamefont {Fedder}}, \bibinfo {author} {\bibfnamefont
  {S.}~\bibnamefont {Paik}}, \bibinfo {author} {\bibfnamefont {L.-P.}\
  \bibnamefont {Yang}}, \bibinfo {author} {\bibfnamefont {N.}~\bibnamefont
  {Zhao}}, \bibinfo {author} {\bibfnamefont {S.}~\bibnamefont {Yang}}, \bibinfo
  {author} {\bibfnamefont {I.}~\bibnamefont {Booker}},  \emph {et~al.},\ }\href
  {\doibase 10.1038/nmat4145} {\bibfield  {journal} {\bibinfo  {journal}
  {Nature materials}\ }\textbf {\bibinfo {volume} {14}},\ \bibinfo {pages}
  {164} (\bibinfo {year} {2015})}\BibitemShut {NoStop}%
\bibitem [{\citenamefont {Christle}\ \emph {et~al.}(2015)\citenamefont
  {Christle}, \citenamefont {Falk}, \citenamefont {Andrich}, \citenamefont
  {Klimov}, \citenamefont {Hassan}, \citenamefont {Son}, \citenamefont
  {Janz\'en}, \citenamefont {Ohshima},\ and\ \citenamefont
  {Awschalom}}]{christle-nature-14}%
  \BibitemOpen
  \bibfield  {author} {\bibinfo {author} {\bibfnamefont {D.~J.}\ \bibnamefont
  {Christle}}, \bibinfo {author} {\bibfnamefont {A.~L.}\ \bibnamefont {Falk}},
  \bibinfo {author} {\bibfnamefont {P.}~\bibnamefont {Andrich}}, \bibinfo
  {author} {\bibfnamefont {P.~V.}\ \bibnamefont {Klimov}}, \bibinfo {author}
  {\bibfnamefont {J.~U.}\ \bibnamefont {Hassan}}, \bibinfo {author}
  {\bibfnamefont {N.~T.}\ \bibnamefont {Son}}, \bibinfo {author} {\bibfnamefont
  {E.}~\bibnamefont {Janz\'en}}, \bibinfo {author} {\bibfnamefont
  {T.}~\bibnamefont {Ohshima}}, \ and\ \bibinfo {author} {\bibfnamefont
  {D.~D.}\ \bibnamefont {Awschalom}},\ }\href
  {https://www.nature.com/articles/nmat4144#supplementary-information}
  {\bibfield  {journal} {\bibinfo  {journal} {Nature materials}\ }\textbf
  {\bibinfo {volume} {14}},\ \bibinfo {pages} {160} (\bibinfo {year}
  {2015})}\BibitemShut {NoStop}%
\bibitem [{\citenamefont {Baranov}\ \emph {et~al.}(2013)\citenamefont
  {Baranov}, \citenamefont {Soltamov}, \citenamefont {Soltamova}, \citenamefont
  {Astakhov},\ and\ \citenamefont {Dyakonov}}]{baranov2013}%
  \BibitemOpen
  \bibfield  {author} {\bibinfo {author} {\bibfnamefont {P.}~\bibnamefont
  {Baranov}}, \bibinfo {author} {\bibfnamefont {V.~A.}\ \bibnamefont
  {Soltamov}}, \bibinfo {author} {\bibfnamefont {A.~A.}\ \bibnamefont
  {Soltamova}}, \bibinfo {author} {\bibfnamefont {G.~V.}\ \bibnamefont
  {Astakhov}}, \ and\ \bibinfo {author} {\bibfnamefont {V.~D.}\ \bibnamefont
  {Dyakonov}},\ }in\ \href {\doibase
  10.4028/www.scientific.net/MSF.740-742.425} {\emph {\bibinfo {booktitle}
  {Silicon Carbide and Related Materials 2012}}},\ \bibinfo {series} {Materials
  Science Forum}, Vol.\ \bibinfo {volume} {740}\ (\bibinfo  {publisher} {Trans
  Tech Publications Ltd},\ \bibinfo {year} {2013})\ pp.\ \bibinfo {pages}
  {425--430}\BibitemShut {NoStop}%
\bibitem [{\citenamefont {Anisimov}\ \emph {et~al.}(2018)\citenamefont
  {Anisimov}, \citenamefont {Soltamov}, \citenamefont {Breev}, \citenamefont
  {Babunts}, \citenamefont {Mokhov}, \citenamefont {Astakhov}, \citenamefont
  {Dyakonov}, \citenamefont {Yakovlev}, \citenamefont {Suter},\ and\
  \citenamefont {Baranov}}]{anisimov-aipa-2018}%
  \BibitemOpen
  \bibfield  {author} {\bibinfo {author} {\bibfnamefont {A.~N.}\ \bibnamefont
  {Anisimov}}, \bibinfo {author} {\bibfnamefont {V.~A.}\ \bibnamefont
  {Soltamov}}, \bibinfo {author} {\bibfnamefont {I.~D.}\ \bibnamefont {Breev}},
  \bibinfo {author} {\bibfnamefont {R.~A.}\ \bibnamefont {Babunts}}, \bibinfo
  {author} {\bibfnamefont {E.~N.}\ \bibnamefont {Mokhov}}, \bibinfo {author}
  {\bibfnamefont {G.~V.}\ \bibnamefont {Astakhov}}, \bibinfo {author}
  {\bibfnamefont {V.}~\bibnamefont {Dyakonov}}, \bibinfo {author}
  {\bibfnamefont {D.~R.}\ \bibnamefont {Yakovlev}}, \bibinfo {author}
  {\bibfnamefont {D.}~\bibnamefont {Suter}}, \ and\ \bibinfo {author}
  {\bibfnamefont {P.~G.}\ \bibnamefont {Baranov}},\ }\href {\doibase
  10.1063/1.5037158} {\bibfield  {journal} {\bibinfo  {journal} {AIP Advances}\
  }\textbf {\bibinfo {volume} {8}},\ \bibinfo {pages} {085304} (\bibinfo {year}
  {2018})}\BibitemShut {NoStop}%
\bibitem [{\citenamefont {Anisimov}\ \emph {et~al.}(2016)\citenamefont
  {Anisimov}, \citenamefont {Simin}, \citenamefont {Soltamov}, \citenamefont
  {Lebedev}, \citenamefont {Baranov}, \citenamefont {Astakhov},\ and\
  \citenamefont {Dyakonov}}]{anisimov-sr-2018}%
  \BibitemOpen
  \bibfield  {author} {\bibinfo {author} {\bibfnamefont {A.}~\bibnamefont
  {Anisimov}}, \bibinfo {author} {\bibfnamefont {D.}~\bibnamefont {Simin}},
  \bibinfo {author} {\bibfnamefont {V.~A.}\ \bibnamefont {Soltamov}}, \bibinfo
  {author} {\bibfnamefont {S.~P.}\ \bibnamefont {Lebedev}}, \bibinfo {author}
  {\bibfnamefont {P.~G.}\ \bibnamefont {Baranov}}, \bibinfo {author}
  {\bibfnamefont {G.~V.}\ \bibnamefont {Astakhov}}, \ and\ \bibinfo {author}
  {\bibfnamefont {V.}~\bibnamefont {Dyakonov}},\ }\href {\doibase
  10.1038/srep33301} {\bibfield  {journal} {\bibinfo  {journal} {Scientific
  reports}\ }\textbf {\bibinfo {volume} {6}},\ \bibinfo {pages} {33301}
  (\bibinfo {year} {2016})}\BibitemShut {NoStop}%
\bibitem [{\citenamefont {Koehl}\ \emph {et~al.}(2011)\citenamefont {Koehl},
  \citenamefont {Buckley}, \citenamefont {Heremans}, \citenamefont {Calusine},\
  and\ \citenamefont {Awschalom}}]{koehl-nature-11}%
  \BibitemOpen
  \bibfield  {author} {\bibinfo {author} {\bibfnamefont {W.~F.}\ \bibnamefont
  {Koehl}}, \bibinfo {author} {\bibfnamefont {B.~B.}\ \bibnamefont {Buckley}},
  \bibinfo {author} {\bibfnamefont {F.~J.}\ \bibnamefont {Heremans}}, \bibinfo
  {author} {\bibfnamefont {G.}~\bibnamefont {Calusine}}, \ and\ \bibinfo
  {author} {\bibfnamefont {D.~D.}\ \bibnamefont {Awschalom}},\ }\href {\doibase
  10.1038/nature10562} {\bibfield  {journal} {\bibinfo  {journal} {Nature}\
  }\textbf {\bibinfo {volume} {479}},\ \bibinfo {pages} {84} (\bibinfo {year}
  {2011})}\BibitemShut {NoStop}%
\bibitem [{\citenamefont {Riedel}\ \emph {et~al.}(2012)\citenamefont {Riedel},
  \citenamefont {Fuchs}, \citenamefont {Kraus}, \citenamefont {V\"ath},
  \citenamefont {Sperlich}, \citenamefont {Dyakonov}, \citenamefont
  {Soltamova}, \citenamefont {Baranov}, \citenamefont {Ilyin},\ and\
  \citenamefont {Astakhov}}]{riedel-prl-12}%
  \BibitemOpen
  \bibfield  {author} {\bibinfo {author} {\bibfnamefont {D.}~\bibnamefont
  {Riedel}}, \bibinfo {author} {\bibfnamefont {F.}~\bibnamefont {Fuchs}},
  \bibinfo {author} {\bibfnamefont {H.}~\bibnamefont {Kraus}}, \bibinfo
  {author} {\bibfnamefont {S.}~\bibnamefont {V\"ath}}, \bibinfo {author}
  {\bibfnamefont {A.}~\bibnamefont {Sperlich}}, \bibinfo {author}
  {\bibfnamefont {V.}~\bibnamefont {Dyakonov}}, \bibinfo {author}
  {\bibfnamefont {A.~A.}\ \bibnamefont {Soltamova}}, \bibinfo {author}
  {\bibfnamefont {P.~G.}\ \bibnamefont {Baranov}}, \bibinfo {author}
  {\bibfnamefont {V.~A.}\ \bibnamefont {Ilyin}}, \ and\ \bibinfo {author}
  {\bibfnamefont {G.~V.}\ \bibnamefont {Astakhov}},\ }\href {\doibase
  10.1103/PhysRevLett.109.226402} {\bibfield  {journal} {\bibinfo  {journal}
  {Physical review letters}\ }\textbf {\bibinfo {volume} {109}},\ \bibinfo
  {pages} {226402} (\bibinfo {year} {2012})}\BibitemShut {NoStop}%
\bibitem [{\citenamefont {Soykal}\ \emph {et~al.}(2016)\citenamefont {Soykal},
  \citenamefont {Dev},\ and\ \citenamefont {Economou}}]{soykal-prb-16}%
  \BibitemOpen
  \bibfield  {author} {\bibinfo {author} {\bibfnamefont {O.~O.}\ \bibnamefont
  {Soykal}}, \bibinfo {author} {\bibfnamefont {P.}~\bibnamefont {Dev}}, \ and\
  \bibinfo {author} {\bibfnamefont {S.~E.}\ \bibnamefont {Economou}},\ }\href
  {https://link.aps.org/doi/10.1103/PhysRevB.93.081207} {\bibfield  {journal}
  {\bibinfo  {journal} {Physical Review B}\ }\textbf {\bibinfo {volume} {93}},\
  \bibinfo {pages} {081207(R)} (\bibinfo {year} {2016})}\BibitemShut {NoStop}%
\bibitem [{\citenamefont {Doherty}\ \emph {et~al.}(2013)\citenamefont
  {Doherty}, \citenamefont {Manson}, \citenamefont {Delaney}, \citenamefont
  {Jelezko}, \citenamefont {Wrachtrup},\ and\ \citenamefont
  {Hollenberg}}]{Doherty:2013uq}%
  \BibitemOpen
  \bibfield  {author} {\bibinfo {author} {\bibfnamefont {M.~W.}\ \bibnamefont
  {Doherty}}, \bibinfo {author} {\bibfnamefont {N.~B.}\ \bibnamefont {Manson}},
  \bibinfo {author} {\bibfnamefont {P.}~\bibnamefont {Delaney}}, \bibinfo
  {author} {\bibfnamefont {F.}~\bibnamefont {Jelezko}}, \bibinfo {author}
  {\bibfnamefont {J.}~\bibnamefont {Wrachtrup}}, \ and\ \bibinfo {author}
  {\bibfnamefont {L.~C.}\ \bibnamefont {Hollenberg}},\ }\href {\doibase
  http://dx.doi.org/10.1016/j.physrep.2013.02.001} {\bibfield  {journal}
  {\bibinfo  {journal} {Physics Reports}\ }\textbf {\bibinfo {volume} {528}},\
  \bibinfo {pages} {1} (\bibinfo {year} {2013})}\BibitemShut {NoStop}%
\bibitem [{\citenamefont {Suter}\ and\ \citenamefont
  {Jelezko}(2017)}]{suter-pnmrs-17}%
  \BibitemOpen
  \bibfield  {author} {\bibinfo {author} {\bibfnamefont {D.}~\bibnamefont
  {Suter}}\ and\ \bibinfo {author} {\bibfnamefont {F.}~\bibnamefont
  {Jelezko}},\ }\href {\doibase 10.1016/j.pnmrs.2016.12.001} {\bibfield
  {journal} {\bibinfo  {journal} {Progress in nuclear magnetic resonance
  spectroscopy}\ }\textbf {\bibinfo {volume} {98}},\ \bibinfo {pages} {50}
  (\bibinfo {year} {2017})}\BibitemShut {NoStop}%
\bibitem [{\citenamefont {Suter}(2020)}]{mr-2020-9}%
  \BibitemOpen
  \bibfield  {author} {\bibinfo {author} {\bibfnamefont {D.}~\bibnamefont
  {Suter}},\ }\href {\doibase 10.5194/mr-1-115-2020} {\bibfield  {journal}
  {\bibinfo  {journal} {Magnetic Resonance}\ }\textbf {\bibinfo {volume} {1}},\
  \bibinfo {pages} {115} (\bibinfo {year} {2020})}\BibitemShut {NoStop}%
\bibitem [{\citenamefont {Falk}\ \emph {et~al.}(2015)\citenamefont {Falk},
  \citenamefont {Klimov}, \citenamefont {Iv\'ady}, \citenamefont {Sz\'asz},
  \citenamefont {Christle}, \citenamefont {Koehl}, \citenamefont {Gali},\ and\
  \citenamefont {Awschalom}}]{falk-prl-15}%
  \BibitemOpen
  \bibfield  {author} {\bibinfo {author} {\bibfnamefont {A.~L.}\ \bibnamefont
  {Falk}}, \bibinfo {author} {\bibfnamefont {P.~V.}\ \bibnamefont {Klimov}},
  \bibinfo {author} {\bibfnamefont {V.}~\bibnamefont {Iv\'ady}}, \bibinfo
  {author} {\bibfnamefont {K.}~\bibnamefont {Sz\'asz}}, \bibinfo {author}
  {\bibfnamefont {D.~J.}\ \bibnamefont {Christle}}, \bibinfo {author}
  {\bibfnamefont {W.~F.}\ \bibnamefont {Koehl}}, \bibinfo {author}
  {\bibfnamefont {A.}~\bibnamefont {Gali}}, \ and\ \bibinfo {author}
  {\bibfnamefont {D.~D.}\ \bibnamefont {Awschalom}},\ }\href {\doibase
  10.1103/PhysRevLett.114.247603} {\bibfield  {journal} {\bibinfo  {journal}
  {Phys. Rev. Lett.}\ }\textbf {\bibinfo {volume} {114}},\ \bibinfo {pages}
  {247603} (\bibinfo {year} {2015})}\BibitemShut {NoStop}%
\bibitem [{\citenamefont {Yan}\ \emph {et~al.}(2018)\citenamefont {Yan},
  \citenamefont {Wang}, \citenamefont {Li}, \citenamefont {Cheng},
  \citenamefont {Cui}, \citenamefont {Liu}, \citenamefont {Xu}, \citenamefont
  {Li},\ and\ \citenamefont {Guo}}]{yan-prap-18}%
  \BibitemOpen
  \bibfield  {author} {\bibinfo {author} {\bibfnamefont {F.-F.}\ \bibnamefont
  {Yan}}, \bibinfo {author} {\bibfnamefont {J.-F.}\ \bibnamefont {Wang}},
  \bibinfo {author} {\bibfnamefont {Q.}~\bibnamefont {Li}}, \bibinfo {author}
  {\bibfnamefont {Z.-D.}\ \bibnamefont {Cheng}}, \bibinfo {author}
  {\bibfnamefont {J.-M.}\ \bibnamefont {Cui}}, \bibinfo {author} {\bibfnamefont
  {W.-Z.}\ \bibnamefont {Liu}}, \bibinfo {author} {\bibfnamefont {J.-S.}\
  \bibnamefont {Xu}}, \bibinfo {author} {\bibfnamefont {C.-F.}\ \bibnamefont
  {Li}}, \ and\ \bibinfo {author} {\bibfnamefont {G.-C.}\ \bibnamefont {Guo}},\
  }\href {\doibase 10.1103/PhysRevApplied.10.044042} {\bibfield  {journal}
  {\bibinfo  {journal} {Phys. Rev. Applied}\ }\textbf {\bibinfo {volume}
  {10}},\ \bibinfo {pages} {044042} (\bibinfo {year} {2018})}\BibitemShut
  {NoStop}%
\bibitem [{\citenamefont {Falk}\ \emph {et~al.}(2014)\citenamefont {Falk},
  \citenamefont {Klimov}, \citenamefont {Buckley}, \citenamefont {Iv\'ady},
  \citenamefont {Abrikosov}, \citenamefont {Calusine}, \citenamefont {Koehl},
  \citenamefont {Gali},\ and\ \citenamefont {Awschalom}}]{falk-prl-14}%
  \BibitemOpen
  \bibfield  {author} {\bibinfo {author} {\bibfnamefont {A.~L.}\ \bibnamefont
  {Falk}}, \bibinfo {author} {\bibfnamefont {P.~V.}\ \bibnamefont {Klimov}},
  \bibinfo {author} {\bibfnamefont {B.~B.}\ \bibnamefont {Buckley}}, \bibinfo
  {author} {\bibfnamefont {V.}~\bibnamefont {Iv\'ady}}, \bibinfo {author}
  {\bibfnamefont {I.~A.}\ \bibnamefont {Abrikosov}}, \bibinfo {author}
  {\bibfnamefont {G.}~\bibnamefont {Calusine}}, \bibinfo {author}
  {\bibfnamefont {W.~F.}\ \bibnamefont {Koehl}}, \bibinfo {author}
  {\bibfnamefont {A.}~\bibnamefont {Gali}}, \ and\ \bibinfo {author}
  {\bibfnamefont {D.~D.}\ \bibnamefont {Awschalom}},\ }\href {\doibase
  10.1103/PhysRevLett.112.187601} {\bibfield  {journal} {\bibinfo  {journal}
  {Phys. Rev. Lett.}\ }\textbf {\bibinfo {volume} {112}},\ \bibinfo {pages}
  {187601} (\bibinfo {year} {2014})}\BibitemShut {NoStop}%
\bibitem [{\citenamefont {Soltamov}\ \emph {et~al.}(2012)\citenamefont
  {Soltamov}, \citenamefont {Soltamova}, \citenamefont {Baranov},\ and\
  \citenamefont {Proskuryakov}}]{soltamov-prl-12}%
  \BibitemOpen
  \bibfield  {author} {\bibinfo {author} {\bibfnamefont {V.~A.}\ \bibnamefont
  {Soltamov}}, \bibinfo {author} {\bibfnamefont {A.~A.}\ \bibnamefont
  {Soltamova}}, \bibinfo {author} {\bibfnamefont {P.~G.}\ \bibnamefont
  {Baranov}}, \ and\ \bibinfo {author} {\bibfnamefont {I.~I.}\ \bibnamefont
  {Proskuryakov}},\ }\href {\doibase 10.1103/PhysRevLett.108.226402} {\bibfield
   {journal} {\bibinfo  {journal} {Phys. Rev. Lett.}\ }\textbf {\bibinfo
  {volume} {108}},\ \bibinfo {pages} {226402} (\bibinfo {year}
  {2012})}\BibitemShut {NoStop}%
\bibitem [{\citenamefont {Biktagirov}\ \emph {et~al.}(2018)\citenamefont
  {Biktagirov}, \citenamefont {Schmidt}, \citenamefont {Gerstmann},
  \citenamefont {Yavkin}, \citenamefont {Orlinskii}, \citenamefont {Baranov},
  \citenamefont {Dyakonov},\ and\ \citenamefont
  {Soltamov}}]{biktagirov-prb-18}%
  \BibitemOpen
  \bibfield  {author} {\bibinfo {author} {\bibfnamefont {T.}~\bibnamefont
  {Biktagirov}}, \bibinfo {author} {\bibfnamefont {W.~G.}\ \bibnamefont
  {Schmidt}}, \bibinfo {author} {\bibfnamefont {U.}~\bibnamefont {Gerstmann}},
  \bibinfo {author} {\bibfnamefont {B.}~\bibnamefont {Yavkin}}, \bibinfo
  {author} {\bibfnamefont {S.}~\bibnamefont {Orlinskii}}, \bibinfo {author}
  {\bibfnamefont {P.}~\bibnamefont {Baranov}}, \bibinfo {author} {\bibfnamefont
  {V.}~\bibnamefont {Dyakonov}}, \ and\ \bibinfo {author} {\bibfnamefont
  {V.}~\bibnamefont {Soltamov}},\ }\href {\doibase 10.1103/PhysRevB.98.195204}
  {\bibfield  {journal} {\bibinfo  {journal} {Phys. Rev. B}\ }\textbf {\bibinfo
  {volume} {98}},\ \bibinfo {pages} {195204} (\bibinfo {year}
  {2018})}\BibitemShut {NoStop}%
\bibitem [{\citenamefont {S\"orman}\ \emph {et~al.}(2000)\citenamefont
  {S\"orman}, \citenamefont {Son}, \citenamefont {Chen}, \citenamefont
  {Kordina}, \citenamefont {Hallin},\ and\ \citenamefont
  {Janz\'en}}]{sorman-prb-00}%
  \BibitemOpen
  \bibfield  {author} {\bibinfo {author} {\bibfnamefont {E.}~\bibnamefont
  {S\"orman}}, \bibinfo {author} {\bibfnamefont {N.~T.}\ \bibnamefont {Son}},
  \bibinfo {author} {\bibfnamefont {W.~M.}\ \bibnamefont {Chen}}, \bibinfo
  {author} {\bibfnamefont {O.}~\bibnamefont {Kordina}}, \bibinfo {author}
  {\bibfnamefont {C.}~\bibnamefont {Hallin}}, \ and\ \bibinfo {author}
  {\bibfnamefont {E.}~\bibnamefont {Janz\'en}},\ }\href {\doibase
  10.1103/PhysRevB.61.2613} {\bibfield  {journal} {\bibinfo  {journal} {Phys.
  Rev. B}\ }\textbf {\bibinfo {volume} {61}},\ \bibinfo {pages} {2613}
  (\bibinfo {year} {2000})}\BibitemShut {NoStop}%
\bibitem [{\citenamefont {Davidsson}\ \emph {et~al.}(2019)\citenamefont
  {Davidsson}, \citenamefont {Ivády}, \citenamefont {Armiento}, \citenamefont
  {Ohshima}, \citenamefont {Son}, \citenamefont {Gali},\ and\ \citenamefont
  {Abrikosov}}]{davidsson-apl-19}%
  \BibitemOpen
  \bibfield  {author} {\bibinfo {author} {\bibfnamefont {J.}~\bibnamefont
  {Davidsson}}, \bibinfo {author} {\bibfnamefont {V.}~\bibnamefont {Ivády}},
  \bibinfo {author} {\bibfnamefont {R.}~\bibnamefont {Armiento}}, \bibinfo
  {author} {\bibfnamefont {T.}~\bibnamefont {Ohshima}}, \bibinfo {author}
  {\bibfnamefont {N.~T.}\ \bibnamefont {Son}}, \bibinfo {author} {\bibfnamefont
  {A.}~\bibnamefont {Gali}}, \ and\ \bibinfo {author} {\bibfnamefont {I.~A.}\
  \bibnamefont {Abrikosov}},\ }\href {\doibase 10.1063/1.5083031} {\bibfield
  {journal} {\bibinfo  {journal} {Applied Physics Letters}\ }\textbf {\bibinfo
  {volume} {114}},\ \bibinfo {pages} {112107} (\bibinfo {year} {2019})},\
  \Eprint {http://arxiv.org/abs/https://doi.org/10.1063/1.5083031}
  {https://doi.org/10.1063/1.5083031} \BibitemShut {NoStop}%
\bibitem [{\citenamefont {Singh}\ \emph {et~al.}(2020)\citenamefont {Singh},
  \citenamefont {Anisimov}, \citenamefont {Nagalyuk}, \citenamefont {Mokhov},
  \citenamefont {Baranov},\ and\ \citenamefont {Suter}}]{singh-prb-20}%
  \BibitemOpen
  \bibfield  {author} {\bibinfo {author} {\bibfnamefont {H.}~\bibnamefont
  {Singh}}, \bibinfo {author} {\bibfnamefont {A.~N.}\ \bibnamefont {Anisimov}},
  \bibinfo {author} {\bibfnamefont {S.~S.}\ \bibnamefont {Nagalyuk}}, \bibinfo
  {author} {\bibfnamefont {E.~N.}\ \bibnamefont {Mokhov}}, \bibinfo {author}
  {\bibfnamefont {P.~G.}\ \bibnamefont {Baranov}}, \ and\ \bibinfo {author}
  {\bibfnamefont {D.}~\bibnamefont {Suter}},\ }\href {\doibase
  10.1103/PhysRevB.101.134110} {\bibfield  {journal} {\bibinfo  {journal}
  {Phys. Rev. B}\ }\textbf {\bibinfo {volume} {101}},\ \bibinfo {pages}
  {134110} (\bibinfo {year} {2020})}\BibitemShut {NoStop}%
\bibitem [{\citenamefont {Nagy}\ \emph {et~al.}(2019)\citenamefont {Nagy},
  \citenamefont {Niethammer}, \citenamefont {Widmann}, \citenamefont {Chen},
  \citenamefont {Udvarhelyi}, \citenamefont {Bonato}, \citenamefont {Hassan},
  \citenamefont {Karhu}, \citenamefont {Ivanov}, \citenamefont {Son} \emph
  {et~al.}}]{nagy-nc-19}%
  \BibitemOpen
  \bibfield  {author} {\bibinfo {author} {\bibfnamefont {R.}~\bibnamefont
  {Nagy}}, \bibinfo {author} {\bibfnamefont {M.}~\bibnamefont {Niethammer}},
  \bibinfo {author} {\bibfnamefont {M.}~\bibnamefont {Widmann}}, \bibinfo
  {author} {\bibfnamefont {Y.-C.}\ \bibnamefont {Chen}}, \bibinfo {author}
  {\bibfnamefont {P.}~\bibnamefont {Udvarhelyi}}, \bibinfo {author}
  {\bibfnamefont {C.}~\bibnamefont {Bonato}}, \bibinfo {author} {\bibfnamefont
  {J.~U.}\ \bibnamefont {Hassan}}, \bibinfo {author} {\bibfnamefont
  {R.}~\bibnamefont {Karhu}}, \bibinfo {author} {\bibfnamefont {I.~G.}\
  \bibnamefont {Ivanov}}, \bibinfo {author} {\bibfnamefont {N.~T.}\
  \bibnamefont {Son}},  \emph {et~al.},\ }\href {\doibase
  10.1038/s41467-019-09873-9} {\bibfield  {journal} {\bibinfo  {journal}
  {Nature communications}\ }\textbf {\bibinfo {volume} {10}},\ \bibinfo {pages}
  {1954} (\bibinfo {year} {2019})}\BibitemShut {NoStop}%
\bibitem [{\citenamefont {Simin}\ \emph {et~al.}(2017)\citenamefont {Simin},
  \citenamefont {Kraus}, \citenamefont {Sperlich}, \citenamefont {Ohshima},
  \citenamefont {Astakhov},\ and\ \citenamefont {Dyakonov}}]{simin-prb-17}%
  \BibitemOpen
  \bibfield  {author} {\bibinfo {author} {\bibfnamefont {D.}~\bibnamefont
  {Simin}}, \bibinfo {author} {\bibfnamefont {H.}~\bibnamefont {Kraus}},
  \bibinfo {author} {\bibfnamefont {A.}~\bibnamefont {Sperlich}}, \bibinfo
  {author} {\bibfnamefont {T.}~\bibnamefont {Ohshima}}, \bibinfo {author}
  {\bibfnamefont {G.~V.}\ \bibnamefont {Astakhov}}, \ and\ \bibinfo {author}
  {\bibfnamefont {V.}~\bibnamefont {Dyakonov}},\ }\href {\doibase
  10.1103/PhysRevB.95.161201} {\bibfield  {journal} {\bibinfo  {journal} {Phys.
  Rev. B}\ }\textbf {\bibinfo {volume} {95}},\ \bibinfo {pages} {161201(R)}
  (\bibinfo {year} {2017})}\BibitemShut {NoStop}%
\bibitem [{\citenamefont {Kraus}\ \emph {et~al.}(2014)\citenamefont {Kraus},
  \citenamefont {Soltamov}, \citenamefont {Riedel}, \citenamefont {V\"ath},
  \citenamefont {Fuchs}, \citenamefont {Sperlich}, \citenamefont {Baranov},
  \citenamefont {Dyakonov},\ and\ \citenamefont {Astakhov}}]{kraus-nature-13}%
  \BibitemOpen
  \bibfield  {author} {\bibinfo {author} {\bibfnamefont {H.}~\bibnamefont
  {Kraus}}, \bibinfo {author} {\bibfnamefont {V.}~\bibnamefont {Soltamov}},
  \bibinfo {author} {\bibfnamefont {D.}~\bibnamefont {Riedel}}, \bibinfo
  {author} {\bibfnamefont {S.}~\bibnamefont {V\"ath}}, \bibinfo {author}
  {\bibfnamefont {F.}~\bibnamefont {Fuchs}}, \bibinfo {author} {\bibfnamefont
  {A.}~\bibnamefont {Sperlich}}, \bibinfo {author} {\bibfnamefont
  {P.}~\bibnamefont {Baranov}}, \bibinfo {author} {\bibfnamefont
  {V.}~\bibnamefont {Dyakonov}}, \ and\ \bibinfo {author} {\bibfnamefont
  {G.}~\bibnamefont {Astakhov}},\ }\href {\doibase 10.1038/nphys2826}
  {\bibfield  {journal} {\bibinfo  {journal} {Nature Physics}\ }\textbf
  {\bibinfo {volume} {10}},\ \bibinfo {pages} {157} (\bibinfo {year}
  {2014})}\BibitemShut {NoStop}%
\bibitem [{\citenamefont {Bathen}\ \emph {et~al.}(2019)\citenamefont {Bathen},
  \citenamefont {Galeckas}, \citenamefont {M\"uting}, \citenamefont {Ayedh},
  \citenamefont {Grossner}, \citenamefont {Coutinho}, \citenamefont
  {Frodason},\ and\ \citenamefont {Vines}}]{bathen-nature-15}%
  \BibitemOpen
  \bibfield  {author} {\bibinfo {author} {\bibfnamefont {M.~E.}\ \bibnamefont
  {Bathen}}, \bibinfo {author} {\bibfnamefont {A.}~\bibnamefont {Galeckas}},
  \bibinfo {author} {\bibfnamefont {J.}~\bibnamefont {M\"uting}}, \bibinfo
  {author} {\bibfnamefont {H.~M.}\ \bibnamefont {Ayedh}}, \bibinfo {author}
  {\bibfnamefont {U.}~\bibnamefont {Grossner}}, \bibinfo {author}
  {\bibfnamefont {J.}~\bibnamefont {Coutinho}}, \bibinfo {author}
  {\bibfnamefont {Y.~K.}\ \bibnamefont {Frodason}}, \ and\ \bibinfo {author}
  {\bibfnamefont {L.}~\bibnamefont {Vines}},\ }\href {\doibase
  10.1038/s41534-019-0227-y} {\bibfield  {journal} {\bibinfo  {journal} {npj
  Quantum Information}\ }\textbf {\bibinfo {volume} {5}},\ \bibinfo {pages} {1}
  (\bibinfo {year} {2019})}\BibitemShut {NoStop}%
\bibitem [{\citenamefont {Soltamov}\ \emph {et~al.}(2019)\citenamefont
  {Soltamov}, \citenamefont {Kasper}, \citenamefont {Poshakinskiy},
  \citenamefont {Anisimov}, \citenamefont {Mokhov}, \citenamefont {Sperlich},
  \citenamefont {Tarasenko}, \citenamefont {Baranov}, \citenamefont
  {Astakhov},\ and\ \citenamefont {Dyakonov}}]{soltamov-naturecom-19}%
  \BibitemOpen
  \bibfield  {author} {\bibinfo {author} {\bibfnamefont {V.}~\bibnamefont
  {Soltamov}}, \bibinfo {author} {\bibfnamefont {C.}~\bibnamefont {Kasper}},
  \bibinfo {author} {\bibfnamefont {A.}~\bibnamefont {Poshakinskiy}}, \bibinfo
  {author} {\bibfnamefont {A.}~\bibnamefont {Anisimov}}, \bibinfo {author}
  {\bibfnamefont {E.}~\bibnamefont {Mokhov}}, \bibinfo {author} {\bibfnamefont
  {A.}~\bibnamefont {Sperlich}}, \bibinfo {author} {\bibfnamefont
  {S.}~\bibnamefont {Tarasenko}}, \bibinfo {author} {\bibfnamefont
  {P.}~\bibnamefont {Baranov}}, \bibinfo {author} {\bibfnamefont
  {G.}~\bibnamefont {Astakhov}}, \ and\ \bibinfo {author} {\bibfnamefont
  {V.}~\bibnamefont {Dyakonov}},\ }\href {\doibase 10.1038/s41467-019-09429-x}
  {\bibfield  {journal} {\bibinfo  {journal} {Nature communications}\ }\textbf
  {\bibinfo {volume} {10}},\ \bibinfo {pages} {1678} (\bibinfo {year}
  {2019})}\BibitemShut {NoStop}%
\bibitem [{\citenamefont {Yang}\ \emph {et~al.}(2014)\citenamefont {Yang},
  \citenamefont {Burk}, \citenamefont {Widmann}, \citenamefont {Lee},
  \citenamefont {Wrachtrup},\ and\ \citenamefont {Zhao}}]{yang-prb-14}%
  \BibitemOpen
  \bibfield  {author} {\bibinfo {author} {\bibfnamefont {L.-P.}\ \bibnamefont
  {Yang}}, \bibinfo {author} {\bibfnamefont {C.}~\bibnamefont {Burk}}, \bibinfo
  {author} {\bibfnamefont {M.}~\bibnamefont {Widmann}}, \bibinfo {author}
  {\bibfnamefont {S.-Y.}\ \bibnamefont {Lee}}, \bibinfo {author} {\bibfnamefont
  {J.}~\bibnamefont {Wrachtrup}}, \ and\ \bibinfo {author} {\bibfnamefont
  {N.}~\bibnamefont {Zhao}},\ }\href {\doibase 10.1103/PhysRevB.90.241203}
  {\bibfield  {journal} {\bibinfo  {journal} {Phys. Rev. B}\ }\textbf {\bibinfo
  {volume} {90}},\ \bibinfo {pages} {241203(R)} (\bibinfo {year}
  {2014})}\BibitemShut {NoStop}%
\bibitem [{\citenamefont {Baranov}\ \emph {et~al.}(2011)\citenamefont
  {Baranov}, \citenamefont {Bundakova}, \citenamefont {Soltamova},
  \citenamefont {Orlinskii}, \citenamefont {Borovykh}, \citenamefont
  {Zondervan}, \citenamefont {Verberk},\ and\ \citenamefont
  {Schmidt}}]{baranov-prb-11}%
  \BibitemOpen
  \bibfield  {author} {\bibinfo {author} {\bibfnamefont {P.~G.}\ \bibnamefont
  {Baranov}}, \bibinfo {author} {\bibfnamefont {A.~P.}\ \bibnamefont
  {Bundakova}}, \bibinfo {author} {\bibfnamefont {A.~A.}\ \bibnamefont
  {Soltamova}}, \bibinfo {author} {\bibfnamefont {S.~B.}\ \bibnamefont
  {Orlinskii}}, \bibinfo {author} {\bibfnamefont {I.~V.}\ \bibnamefont
  {Borovykh}}, \bibinfo {author} {\bibfnamefont {R.}~\bibnamefont {Zondervan}},
  \bibinfo {author} {\bibfnamefont {R.}~\bibnamefont {Verberk}}, \ and\
  \bibinfo {author} {\bibfnamefont {J.}~\bibnamefont {Schmidt}},\ }\href
  {\doibase 10.1103/PhysRevB.83.125203} {\bibfield  {journal} {\bibinfo
  {journal} {Phys. Rev. B}\ }\textbf {\bibinfo {volume} {83}},\ \bibinfo
  {pages} {125203} (\bibinfo {year} {2011})}\BibitemShut {NoStop}%
\bibitem [{\citenamefont {Fuchs}\ \emph {et~al.}(2015)\citenamefont {Fuchs},
  \citenamefont {Stender}, \citenamefont {Trupke}, \citenamefont {Simin},
  \citenamefont {Pflaum}, \citenamefont {Dyakonov},\ and\ \citenamefont
  {Astakhov}}]{fuchs-nature-15}%
  \BibitemOpen
  \bibfield  {author} {\bibinfo {author} {\bibfnamefont {F.}~\bibnamefont
  {Fuchs}}, \bibinfo {author} {\bibfnamefont {B.}~\bibnamefont {Stender}},
  \bibinfo {author} {\bibfnamefont {M.}~\bibnamefont {Trupke}}, \bibinfo
  {author} {\bibfnamefont {D.}~\bibnamefont {Simin}}, \bibinfo {author}
  {\bibfnamefont {J.}~\bibnamefont {Pflaum}}, \bibinfo {author} {\bibfnamefont
  {V.}~\bibnamefont {Dyakonov}}, \ and\ \bibinfo {author} {\bibfnamefont
  {G.}~\bibnamefont {Astakhov}},\ }\href {\doibase 10.1038/ncomms8578}
  {\bibfield  {journal} {\bibinfo  {journal} {Nature communications}\ }\textbf
  {\bibinfo {volume} {6}},\ \bibinfo {pages} {7578} (\bibinfo {year}
  {2015})}\BibitemShut {NoStop}%
\bibitem [{\citenamefont {Astakhov}\ \emph {et~al.}(2016)\citenamefont
  {Astakhov}, \citenamefont {Simin}, \citenamefont {Dyakonov}, \citenamefont
  {Yavkin}, \citenamefont {Orlinskii}, \citenamefont {Proskuryakov},
  \citenamefont {Anisimov}, \citenamefont {Soltamov},\ and\ \citenamefont
  {Baranov}}]{astakhov2016spin}%
  \BibitemOpen
  \bibfield  {author} {\bibinfo {author} {\bibfnamefont {G.}~\bibnamefont
  {Astakhov}}, \bibinfo {author} {\bibfnamefont {D.}~\bibnamefont {Simin}},
  \bibinfo {author} {\bibfnamefont {V.}~\bibnamefont {Dyakonov}}, \bibinfo
  {author} {\bibfnamefont {B.}~\bibnamefont {Yavkin}}, \bibinfo {author}
  {\bibfnamefont {S.}~\bibnamefont {Orlinskii}}, \bibinfo {author}
  {\bibfnamefont {I.}~\bibnamefont {Proskuryakov}}, \bibinfo {author}
  {\bibfnamefont {A.}~\bibnamefont {Anisimov}}, \bibinfo {author}
  {\bibfnamefont {V.}~\bibnamefont {Soltamov}}, \ and\ \bibinfo {author}
  {\bibfnamefont {P.}~\bibnamefont {Baranov}},\ }\href {\doibase
  10.1007/s00723-016-0800-x} {\bibfield  {journal} {\bibinfo  {journal}
  {Applied Magnetic Resonance}\ }\textbf {\bibinfo {volume} {47}},\ \bibinfo
  {pages} {793} (\bibinfo {year} {2016})}\BibitemShut {NoStop}%
\bibitem [{\citenamefont {Hain}\ \emph {et~al.}(2014)\citenamefont {Hain},
  \citenamefont {Fuchs}, \citenamefont {Soltamov}, \citenamefont {Baranov},
  \citenamefont {Astakhov}, \citenamefont {Hertel},\ and\ \citenamefont
  {Dyakonov}}]{hain-jap-14}%
  \BibitemOpen
  \bibfield  {author} {\bibinfo {author} {\bibfnamefont {T.~C.}\ \bibnamefont
  {Hain}}, \bibinfo {author} {\bibfnamefont {F.}~\bibnamefont {Fuchs}},
  \bibinfo {author} {\bibfnamefont {V.~A.}\ \bibnamefont {Soltamov}}, \bibinfo
  {author} {\bibfnamefont {P.~G.}\ \bibnamefont {Baranov}}, \bibinfo {author}
  {\bibfnamefont {G.~V.}\ \bibnamefont {Astakhov}}, \bibinfo {author}
  {\bibfnamefont {T.}~\bibnamefont {Hertel}}, \ and\ \bibinfo {author}
  {\bibfnamefont {V.}~\bibnamefont {Dyakonov}},\ }\href {\doibase
  10.1063/1.4870456} {\bibfield  {journal} {\bibinfo  {journal} {Journal of
  Applied Physics}\ }\textbf {\bibinfo {volume} {115}},\ \bibinfo {pages}
  {133508} (\bibinfo {year} {2014})}\BibitemShut {NoStop}%
\bibitem [{\citenamefont {Chen}(2003)}]{Chen2003}%
  \BibitemOpen
  \bibfield  {author} {\bibinfo {author} {\bibfnamefont {W.~M.}\ \bibnamefont
  {Chen}},\ }\enquote {\bibinfo {title} {Optically detected magnetic resonance
  of defects in semiconductors},}\ in\ \href {\doibase
  10.1007/978-1-4757-5166-6_15} {\emph {\bibinfo {booktitle} {EPR of Free
  Radicals in Solids: Trends in Methods and Applications}}},\ \bibinfo {editor}
  {edited by\ \bibinfo {editor} {\bibfnamefont {A.}~\bibnamefont {Lund}}\ and\
  \bibinfo {editor} {\bibfnamefont {M.}~\bibnamefont {Shiotani}}}\ (\bibinfo
  {publisher} {Springer US},\ \bibinfo {address} {Boston, MA},\ \bibinfo {year}
  {2003})\ pp.\ \bibinfo {pages} {601--625}\BibitemShut {NoStop}%
\bibitem [{\citenamefont {Depinna}\ and\ \citenamefont
  {Cavenett}(1982)}]{Depinna-1982}%
  \BibitemOpen
  \bibfield  {author} {\bibinfo {author} {\bibfnamefont {S.}~\bibnamefont
  {Depinna}}\ and\ \bibinfo {author} {\bibfnamefont {B.}~\bibnamefont
  {Cavenett}},\ }\href@noop {} {\bibfield  {journal} {\bibinfo  {journal}
  {Journal of Physics C: Solid State Physics}\ }\textbf {\bibinfo {volume}
  {15}},\ \bibinfo {pages} {L489} (\bibinfo {year} {1982})}\BibitemShut
  {NoStop}%
\bibitem [{\citenamefont {Langof}\ \emph {et~al.}(2002)\citenamefont {Langof},
  \citenamefont {Ehrenfreund}, \citenamefont {Lifshitz}, \citenamefont
  {Micic},\ and\ \citenamefont {Nozik}}]{langof-jpcb-02}%
  \BibitemOpen
  \bibfield  {author} {\bibinfo {author} {\bibfnamefont {L.}~\bibnamefont
  {Langof}}, \bibinfo {author} {\bibfnamefont {E.}~\bibnamefont {Ehrenfreund}},
  \bibinfo {author} {\bibfnamefont {E.}~\bibnamefont {Lifshitz}}, \bibinfo
  {author} {\bibfnamefont {O.~I.}\ \bibnamefont {Micic}}, \ and\ \bibinfo
  {author} {\bibfnamefont {A.~J.}\ \bibnamefont {Nozik}},\ }\href {\doibase
  10.1021/jp013720g} {\bibfield  {journal} {\bibinfo  {journal} {The Journal of
  Physical Chemistry B}\ }\textbf {\bibinfo {volume} {106}},\ \bibinfo {pages}
  {1606} (\bibinfo {year} {2002})}\BibitemShut {NoStop}%
\bibitem [{\citenamefont {Carter}\ \emph {et~al.}(2015)\citenamefont {Carter},
  \citenamefont {Soykal}, \citenamefont {Dev}, \citenamefont {Economou},\ and\
  \citenamefont {Glaser}}]{carter-prb-15}%
  \BibitemOpen
  \bibfield  {author} {\bibinfo {author} {\bibfnamefont {S.~G.}\ \bibnamefont
  {Carter}}, \bibinfo {author} {\bibfnamefont {O.~O.}\ \bibnamefont {Soykal}},
  \bibinfo {author} {\bibfnamefont {P.}~\bibnamefont {Dev}}, \bibinfo {author}
  {\bibfnamefont {S.~E.}\ \bibnamefont {Economou}}, \ and\ \bibinfo {author}
  {\bibfnamefont {E.~R.}\ \bibnamefont {Glaser}},\ }\href {\doibase
  10.1103/PhysRevB.92.161202} {\bibfield  {journal} {\bibinfo  {journal} {Phys.
  Rev. B}\ }\textbf {\bibinfo {volume} {92}},\ \bibinfo {pages} {161202(R)}
  (\bibinfo {year} {2015})}\BibitemShut {NoStop}%
\bibitem [{\citenamefont {Ramsey}(1950)}]{ramsey-pr-50}%
  \BibitemOpen
  \bibfield  {author} {\bibinfo {author} {\bibfnamefont {N.~F.}\ \bibnamefont
  {Ramsey}},\ }\href {\doibase 10.1103/PhysRev.78.695} {\bibfield  {journal}
  {\bibinfo  {journal} {Phys. Rev.}\ }\textbf {\bibinfo {volume} {78}},\
  \bibinfo {pages} {695} (\bibinfo {year} {1950})}\BibitemShut {NoStop}%
\bibitem [{\citenamefont {Eickhoff}\ and\ \citenamefont
  {Suter}(2004)}]{EICKHOFF200469}%
  \BibitemOpen
  \bibfield  {author} {\bibinfo {author} {\bibfnamefont {M.}~\bibnamefont
  {Eickhoff}}\ and\ \bibinfo {author} {\bibfnamefont {D.}~\bibnamefont
  {Suter}},\ }\href {\doibase https://doi.org/10.1016/j.jmr.2003.09.009}
  {\bibfield  {journal} {\bibinfo  {journal} {Journal of Magnetic Resonance}\
  }\textbf {\bibinfo {volume} {166}},\ \bibinfo {pages} {69 } (\bibinfo {year}
  {2004})}\BibitemShut {NoStop}%
\bibitem [{\citenamefont {Hahn}(1950)}]{hahn-pr-50}%
  \BibitemOpen
  \bibfield  {author} {\bibinfo {author} {\bibfnamefont {E.~L.}\ \bibnamefont
  {Hahn}},\ }\href {\doibase 10.1103/PhysRev.80.580} {\bibfield  {journal}
  {\bibinfo  {journal} {Phys. Rev.}\ }\textbf {\bibinfo {volume} {80}},\
  \bibinfo {pages} {580} (\bibinfo {year} {1950})}\BibitemShut {NoStop}%
\bibitem [{\citenamefont {Bloembergen}\ \emph {et~al.}(1948)\citenamefont
  {Bloembergen}, \citenamefont {Purcell},\ and\ \citenamefont {Pound}}]{550}%
  \BibitemOpen
  \bibfield  {author} {\bibinfo {author} {\bibfnamefont {N.}~\bibnamefont
  {Bloembergen}}, \bibinfo {author} {\bibfnamefont {E.~M.}\ \bibnamefont
  {Purcell}}, \ and\ \bibinfo {author} {\bibfnamefont {R.~V.}\ \bibnamefont
  {Pound}},\ }\href {\doibase 10.1103/PhysRev.73.679} {\bibfield  {journal}
  {\bibinfo  {journal} {Phys. Rev.}\ }\textbf {\bibinfo {volume} {73}},\
  \bibinfo {pages} {679} (\bibinfo {year} {1948})}\BibitemShut {NoStop}%
\bibitem [{\citenamefont {Abragam}(1961)}]{abragam-book}%
  \BibitemOpen
  \bibfield  {author} {\bibinfo {author} {\bibfnamefont {A.}~\bibnamefont
  {Abragam}},\ }\href@noop {} {\emph {\bibinfo {title} {The principles of
  nuclear magnetism}}}\ (\bibinfo  {publisher} {Oxford University Press},\
  \bibinfo {address} {UK},\ \bibinfo {year} {1961})\BibitemShut {NoStop}%
\bibitem [{\citenamefont {DiVincenzo}(2000)}]{divincenzo}%
  \BibitemOpen
  \bibfield  {author} {\bibinfo {author} {\bibfnamefont {D.~P.}\ \bibnamefont
  {DiVincenzo}},\ }\href {\doibase
  10.1002/1521-3978(200009)48:9/11<771::AID-PROP771>3.0.CO;2-E} {\bibfield
  {journal} {\bibinfo  {journal} {Fortschritte der Physik: Progress of
  Physics}\ }\textbf {\bibinfo {volume} {48}},\ \bibinfo {pages} {771}
  (\bibinfo {year} {2000})}\BibitemShut {NoStop}%
\bibitem [{\citenamefont {Stolze}\ and\ \citenamefont
  {Suter}(2008)}]{Stolze:2008xy}%
  \BibitemOpen
  \bibfield  {author} {\bibinfo {author} {\bibfnamefont {J.}~\bibnamefont
  {Stolze}}\ and\ \bibinfo {author} {\bibfnamefont {D.}~\bibnamefont {Suter}},\
  }\href@noop {} {\emph {\bibinfo {title} {Quantum Computing: A Short Course
  from Theory to Experiment}}},\ \bibinfo {edition} {2nd}\ ed.\ (\bibinfo
  {publisher} {Wiley-VCH},\ \bibinfo {address} {Berlin},\ \bibinfo {year}
  {2008})\BibitemShut {NoStop}%
\bibitem [{\citenamefont {Tarasenko}\ \emph {et~al.}(2018)\citenamefont
  {Tarasenko}, \citenamefont {Poshakinskiy}, \citenamefont {Simin},
  \citenamefont {Soltamov}, \citenamefont {Mokhov}, \citenamefont {Baranov},
  \citenamefont {Dyakonov},\ and\ \citenamefont
  {Astakhov}}]{tarasenko-pssb-18}%
  \BibitemOpen
  \bibfield  {author} {\bibinfo {author} {\bibfnamefont {S.~A.}\ \bibnamefont
  {Tarasenko}}, \bibinfo {author} {\bibfnamefont {A.~V.}\ \bibnamefont
  {Poshakinskiy}}, \bibinfo {author} {\bibfnamefont {D.}~\bibnamefont {Simin}},
  \bibinfo {author} {\bibfnamefont {V.~A.}\ \bibnamefont {Soltamov}}, \bibinfo
  {author} {\bibfnamefont {E.~N.}\ \bibnamefont {Mokhov}}, \bibinfo {author}
  {\bibfnamefont {P.~G.}\ \bibnamefont {Baranov}}, \bibinfo {author}
  {\bibfnamefont {V.}~\bibnamefont {Dyakonov}}, \ and\ \bibinfo {author}
  {\bibfnamefont {G.~V.}\ \bibnamefont {Astakhov}},\ }\href {\doibase
  https://doi.org/10.1002/pssb.201870101} {\bibfield  {journal} {\bibinfo
  {journal} {physica status solidi (b)}\ }\textbf {\bibinfo {volume} {255}},\
  \bibinfo {pages} {1870101} (\bibinfo {year} {2018})},\ \Eprint
  {http://arxiv.org/abs/https://onlinelibrary.wiley.com/doi/pdf/10.1002/pssb.201870101}
  {https://onlinelibrary.wiley.com/doi/pdf/10.1002/pssb.201870101} \BibitemShut
  {NoStop}%
\bibitem [{\citenamefont {Ramsay}\ and\ \citenamefont
  {Rossi}(2020)}]{ramsay-prb-20}%
  \BibitemOpen
  \bibfield  {author} {\bibinfo {author} {\bibfnamefont {A.~J.}\ \bibnamefont
  {Ramsay}}\ and\ \bibinfo {author} {\bibfnamefont {A.}~\bibnamefont {Rossi}},\
  }\href {\doibase 10.1103/PhysRevB.101.165307} {\bibfield  {journal} {\bibinfo
   {journal} {Phys. Rev. B}\ }\textbf {\bibinfo {volume} {101}},\ \bibinfo
  {pages} {165307} (\bibinfo {year} {2020})}\BibitemShut {NoStop}%
\bibitem [{\citenamefont {van~der Maarel}(2003)}]{johan-cmr-03}%
  \BibitemOpen
  \bibfield  {author} {\bibinfo {author} {\bibfnamefont {J.~R.}\ \bibnamefont
  {van~der Maarel}},\ }\href {\doibase https://doi.org/10.1002/cmr.a.10087}
  {\bibfield  {journal} {\bibinfo  {journal} {Concepts in Magnetic Resonance
  Part A}\ }\textbf {\bibinfo {volume} {19A}},\ \bibinfo {pages} {97} (\bibinfo
  {year} {2003})},\ \Eprint
  {http://arxiv.org/abs/https://onlinelibrary.wiley.com/doi/pdf/10.1002/cmr.a.10087}
  {https://onlinelibrary.wiley.com/doi/pdf/10.1002/cmr.a.10087} \BibitemShut
  {NoStop}%
\bibitem [{\citenamefont {Mizuochi}\ \emph {et~al.}(2002)\citenamefont
  {Mizuochi}, \citenamefont {Yamasaki}, \citenamefont {Takizawa}, \citenamefont
  {Morishita}, \citenamefont {Ohshima}, \citenamefont {Itoh},\ and\
  \citenamefont {Isoya}}]{Mizuochi-prb-02}%
  \BibitemOpen
  \bibfield  {author} {\bibinfo {author} {\bibfnamefont {N.}~\bibnamefont
  {Mizuochi}}, \bibinfo {author} {\bibfnamefont {S.}~\bibnamefont {Yamasaki}},
  \bibinfo {author} {\bibfnamefont {H.}~\bibnamefont {Takizawa}}, \bibinfo
  {author} {\bibfnamefont {N.}~\bibnamefont {Morishita}}, \bibinfo {author}
  {\bibfnamefont {T.}~\bibnamefont {Ohshima}}, \bibinfo {author} {\bibfnamefont
  {H.}~\bibnamefont {Itoh}}, \ and\ \bibinfo {author} {\bibfnamefont
  {J.}~\bibnamefont {Isoya}},\ }\href {\doibase 10.1103/PhysRevB.66.235202}
  {\bibfield  {journal} {\bibinfo  {journal} {Phys. Rev. B}\ }\textbf {\bibinfo
  {volume} {66}},\ \bibinfo {pages} {235202} (\bibinfo {year}
  {2002})}\BibitemShut {NoStop}%
\end{thebibliography}
\end{document}